\documentclass[11pt]{article}
\usepackage{amssymb,color}
\newcommand{\N}{N\raise.7ex\hbox{\underline{$\circ $}}$\;$}

\begin{document}

\title{General conditions for vanishing the current  $J^{z}$ for  Dirac field
on  boundaries of the domain between two planes
 }

\maketitle

\author{O.V. Veko\footnote{Kalinkovichi Gymnasium,
Belarus,vekoolga@mail.ru}, V.M. Red'kov\footnote{B.I. Stepanov
Institute of Physics, NAS of Belarus, redkov@dragon.bas-net.by},
A.I. Shelest\footnote{B.I. Stepanov Institute of Physics, NAS of
Belarus, afanasie@tut.by}, S.A. Yushchenko\footnote{N.N.
Bogolyubov Institute for Theoretical Physics. NAS of Ukraine,
yushchenko@bitp.kiev.ua}, A.M. Ishkhanyan\footnote{ Institute for
Physical Research, Armenian Academy of Sciences,
aishkhanyan@gmail.com}}

\begin{abstract}

In connection with the Casimir effect for a   spinor   field in
presence of external magnetic field, of special interest are
solutions of the Dirac equation in the domains restricted by two
planes, which have vanishing the third projection of the conserved
current $J^{z}$ on two boundaries. General conditions for
vanishing the current  are formulated, they  reduce to linear
homogeneous algebraic system, for which solutions exist
only when vanishing the determinant of the linear system, that is
for the roots of  a 4-th
 order algebraic equation  with respect to the variable
 $e^{2ik a}$,  where $a$ is a half-distance between the planes, and $k$ stands for
 the third projection of the Dirac  particle momentum.
All  solutions  of the equation  have been found explicitly, each of them provides us in principle with a
special possibility to get the quantization rules for parameter $k$; the most of produced expression for the roots
can be solved with respect to parameter $k$ only numerically. Generally, solutions $e^{2ik a}$  depend
on  4 arbitrary phase parameters which influence the  appropriate
wave functions
 with vanishing current:
$J^{z}(z=-z,+a)=0$.

\end{abstract}

{\bf PACS numbers}: 02.30.Gp, 02.40.Ky, 03.65Ge, 04.62.+v

{\bf MSC 2010:} 33E30, 34B30.\\

{\bf Keywords} Casimir effect, Dirac field, magnetic field,
conserved current, quantization, 4-th order polynomial

\section{Introduction}

The  boundary conditions imposed on a quantum field leads to the
modification of the vacuum energy levels and can be observed experimentally
as a  vacuum pressure. This effect (called Casimir effect) has been predicted
by Casimir in his original work in 1948 \cite{Casimir-1948}, \cite{Casimir-1951},
 and has been experimentally
observed for electromagnetic field several years later. Until now, in many
theoretical works Casimir energy has been computed for various types of
boundary geometry and for fields other than electromagnetic one.
It  has received much attention since its discovery -- see review
\cite{Mostenanenko-Trunov-1988}, \cite{Mostepanenk-Trunov-1997}, and more recent
\cite{Bordag-Klimchitskaya-Mohideen-Mostepanenko}.
The Casimir effect  for charged relativistic fields can be also influenced by
external electromagnetic  fields.

In contrast to boson fields (for instance, scalar or electromagnetic), for  the Dirac  field
one cannot impose directly restriction of  vanishing the field on the boundary of the restricted domain.
In 1975, Johnson \cite{Johnson-1975}  calculated for the first time the Casimir fermionic
effect in the MIT-bag model. Johnson used as restriction the condition of vanishing the
  normal projection of the
current of the Dirac field at the boundary surface of a domain (the bag).

The perturbation of Casimir energy by an external magnetic field was
firstly considered by Elizalde et al
 \cite{Elizalde-et-al-1997}  and Cougo-Pinto et al \cite{Cougo-Pinto-et-al-1999}.
  The influence on the Casimir energy of the Dirac field was firstly studied for an
 antiperiodic boundary condition by Farina -- Tort --  Cougo--Pinto
\cite{Farina-Tort-Cougo-Pinto}.
 These authors showed
that the Casimir effect can be enhanced by a magnetic field.

 More recently the influence of the external magnetic field on the
complex scalar field as well as on the Casimir energy of the Dirac field was
considered by Ostrowski \cite{Ostrowski} who calculates these by direct solving the field
equations using the mode summation method. He has considered  a complex scalar field
 confined between two infinite plates with Dirichlet boundary conditions with magnetic
field in the  direction perpendicular to the plates. Also he  considered
the  Dirac field with  antiperiodic boundary conditions.
In scalar case the most simple and popular is the  Dirichlet boundary conditions in the form:
$$
\Phi(x,y,z=-a) = \Phi(x,y,z=+a)
$$
which implies that field  is equal to zero on the two parallel planes (plates).
The plates are perpendicular to the $z$-axis and the distance between them
is equal to $2a$.
In the fermionic case the most popular is
 antiperiodic
boundary condition in  the form:
$$
\Psi (x,y,z+2a) = -\Psi (x,y,z)\; .
$$
Finally in the the recent study of the Casimir effect for charged scalar field in presence of
magnetic
field
\cite{Sitenko-Yushchenko},  it is shown that in the case
of a weak magnetic field and a small separation of the plates, the
Casimir force is either attractive or repulsive, depending on the choice
of a boundary condition. In the case of a strong magnetic field and
a large separation of the plates, the Casimir force is repulsive, being
independent of the choice of a boundary condition, as well as of the
distance between the plates.

In the present paper we turn back to the case of Dirac field bounded by two parallel planes
and submitted to
 external uniform magnetic field;  the aim is to study in full  generality
the Jhonson  MIT-bag condition for the current $J^{z}$ on the boundaries of the domain between two planes.
As demonstrates the present analysis, the known in the literature associated restriction
is (see for instance in  \cite{Mostenanenko-Trunov-1988}) only simplest and particular example, whereas
 many other  possibilities exist  here as well.

The outline of the paper is the following.
In connection with the Casimir effect for the Dirac field in
external  magnetic field, of special interest are
solutions of the Dirac equation in the domains between  two
planes, which have vanishing third projection of the conserved
current $J^{z}$ on the  boundaries. Solutions with  such
properties are reachable when considering 4-dimensional linear
space $\{\Psi \}$  based of four solutions for Dirac particle  with opposite signs of the third
projection of momentum
 $+k$ and  $-k$
  along the direction of the magnetic field.
General conditions for vanishing such a current at the boundaries
are formulated, they  reduce to linear homogeneous algebraic
system, for which equation $\det S =0$ is a 4-th
 order algebraic equation, the roots $K$
of which are
 $K= e^{2ik a}$,  where $a$ is a half-distance between the planes, and $k$ stands for
 the third projection of the Dirac  particle momentum.
Each root  of this  equation , if it   represents  a complex number
of the unit length, will give  a certain rule for quantization of
the third projection of $k$. The complete analysis and classification
of all such solutions has been performed.

\section{ Solutions of he Dirac equation in magnetic field
}

Let us consider the  Dirac equation in Cartesian coordinates in
presence of external uniform magnetic field  ${\bf A} = ( -  B  y,
0,0)$
\begin{eqnarray}
 [ \gamma^{0}i{\partial \over \partial t} + \gamma^{1} (  i
{\partial \over \partial x} -  B y  )   \nonumber \\
+  i \gamma^{2} {\partial \over \partial y}
 + i \gamma^{3} {\partial \over \partial z}  -M
 ] \Psi=0. \label{2.2}
\end{eqnarray}
Below we use the spinor basis for the Dirac matrices and the
following substitution for the wave functions
\begin{eqnarray}
 \Psi = e^{-i\epsilon t} e^{iax} e^{ikz} \left |
\begin{array}{c}
f_{1}(y) \\
f_{2}(y) \\
f_{3}(y) \\
f_{4}(y) \\
\end{array} \right |;
\label{2.3}
\end{eqnarray}

\noindent which results in equations for  $f_{a}(y)$
\begin{eqnarray}
\epsilon f_{3}+ (a+By)f_{4}- {\partial \over \partial y}f_{4}
+kf_{3} -Mf_{1}= 0\; ,
\nonumber
\\
\epsilon f_{4}+ (a+By)f_{3}+ {\partial \over \partial
y}f_{3}-kf_{4} -Mf_{2}= 0\; ,
\nonumber
\\
\epsilon f_{1}- (a+By)f_{2} +{\partial \over \partial
y}f_{2}-kf_{1} -Mf_{3}= 0\; ,
\nonumber
\\
\epsilon f_{2}- (a+By)f_{1}-{\partial \over \partial
y}f_{1}+kf_{2} -Mf_{4}= 0\; . \label{2.4}
\end{eqnarray}

\noindent
 Let us impose linear restrictions (below we will prove
their consistency)
\begin{eqnarray}
f_{3} = A f_{1}, \qquad  f_{4} = A f_{2}\, ; \label{2.5}
\end{eqnarray}

\noindent
system  (\ref{2.4}) will take the form
\begin{eqnarray}
(\epsilon  +k - {M \over A} )f_{1}+ [(a+By)- {\partial \over
\partial y}]  f_{2}  = 0\; ,
\nonumber \\
 (-\epsilon  + k +{M \over A}) f_{2} - [(a+ By) +
{\partial \over
\partial y}]  f_{1}  = 0\; ,
\nonumber
\\
(-\epsilon + k +MA) f_{1} +[ (a+By) -{\partial \over \partial
y}]f_{2}  = 0\; , \nonumber \\
 (\epsilon + k -MA) f_{2}-[ (a+By)
+{\partial \over \partial y } ]f_{1} = 0\;.  \label{2.6}
\end{eqnarray}

\noindent Equations 1 and 3, as well as  2 and 4,  will coincide,
if  the condition below holds (we will use notation
$+\sqrt{\epsilon^{2}-M^{2}}=p$)
\begin{eqnarray}
 A = { \epsilon\pm p  \over M} \,.
 \label{2.7}
 \end{eqnarray}

\noindent Depending on values of  $A$, we have two cases:

\vspace{2mm} $ A = {\epsilon + p \over M} = \alpha , $
\begin{eqnarray}
(  k +p) f_{1} +  [  (a+ B y) -{\partial \over \partial y}  ]
f_{2}  = 0\; , \nonumber \\
 ( k -p) f_{2}-  [  (a+ B y) +{\partial
\over
\partial y }  ] f_{1} = 0\; ; \label{2.8a}
\end{eqnarray}

$ A = {\epsilon - p \over M} = \beta,$
\begin{eqnarray}
( k -p) f_{1} +[ (a+ B y) -{\partial \over \partial y}]f_{2}  =
0\; , \nonumber \\
 ( k +p) f_{2}-[ (a+ B y) +{\partial \over
\partial y } ]f_{1} = 0\; . \label{2.8b}
\end{eqnarray}

Let us specify explicit form of an operator related to
linear restriction  (\ref{2.5}). For a free Dirac particle, that is
 the helicity operator
\begin{eqnarray}
\Sigma_{0}  =  \vec{\Sigma} \; \vec{P} = {1 \over 2}  \left (
\gamma^{2}\gamma^{3} {\partial \over \partial x } +
 \gamma^{3}\gamma^{1}  {\partial \over \partial y }  +
 \gamma^{1}\gamma^{2}  {\partial  \over \partial z }   \right ).
\nonumber
\end{eqnarray}
In presence of the  magnetic field, it becomes more complicated
\begin{eqnarray}
\Sigma  =    {1 \over 2}  \left (  \gamma^{2}\gamma^{3} (
{\partial \over \partial x } + i B y )  +
 \gamma^{3}\gamma^{1}  {\partial \over \partial y }  +
 \gamma^{1}\gamma^{2}  {\partial  \over \partial z }   \right ).
\nonumber
\label{2.9}
\end{eqnarray}

\noindent The eigenvalue  equation
 $\Sigma \Psi = p \Psi $ gives
\begin{eqnarray}
( a + B y) f_{2} \; - \;  {d \over dy }  f_{2} \; +  k f_{1} =  p
f_{1} \; , \nonumber \\
 (a + B y) f_{1} \; +\;   {d \over dy }
f_{1} \; - k f_{2} =  p f_{2} \; , \nonumber
\\
(a + B y)  f_{4} \; - \;  {d \over dy }  f_{4} \; +  k \; f_{3} =
p f_{3} \; , \nonumber \\  (a + B y) f_{3} \; + \;  {d \over dy}
f_{3} \; - k  f_{4} =  p f_{4} \;.
 \nonumber
 \end{eqnarray}

\noindent Considering them jointly with   (\ref{2.4}), we arrive
at the linear algebraic system
\begin{eqnarray}
 p  \, f_{3}
+  ( \epsilon   f_{3}   -  M  f_{1} ) =0\, ,\;
 p \,
f_{4}
 +    (  \epsilon  f_{4}   - M  f_{2} ) =0\, ,
\nonumber
\\
 p \, f_{1}
 -  ( \epsilon f_{1} -M f_{3} ) =0\, ,
 \;p  \, f_{2}
   - (   \epsilon f_{2} -M f_{4} ) =0\,  ,
\nonumber
\end{eqnarray}

\noindent from whence it follows
\begin{eqnarray}
p =  + \sqrt{\epsilon^{2} - m^{2}} ,\; f_{3} = {\epsilon \pm  p
\over M} f_{1}  ,\;
 f_{4} = {\epsilon  \pm  p  \over M} f_{2}  ;
\nonumber
\end{eqnarray}

\noindent these coincide with the above used restrictions.

Remembering on this, let us write down the substitutions for the wave
functions of opposite polarization states:
\begin{eqnarray}
 \Psi_{\epsilon a k, \alpha} =e^{-i\epsilon t} e^{iax}
e^{ikz} \left | \begin{array}{r}
f_{1}(y)\\
f_{2}(y)\\
\alpha f_{1}(y)\\
\alpha f_{2}(y) \
\end{array} \right |,
\nonumber
\\
 \Psi_{\epsilon a k, \beta} =e^{-i\epsilon t} e^{iax}
e^{ikz} \left | \begin{array}{r}
f_{1}(y)\\
f_{2}(y)\\
\beta f_{1}(y)\\
\beta f_{2}(y) \
\end{array} \right |;
\label{2.10}
\end{eqnarray}

\noindent these waves should be solutions of respective systems:
(note the notation: $\lambda^{2} =  \epsilon^{2} - m^{2} -k^{2}$)

type-$\alpha$,
\begin{eqnarray}
{d^{2}f_{1}\over dy^{2}}+\left[ B -( a+ B
y)^{2}+\lambda^{2}\right]f_{1}=0\,, \nonumber \\
 f_{2} = {1 \over
k -p} \left ( {d \over dy} + (a+ B y) \right ) f_{1} \; ;
\nonumber
\label{2.11a}
\end{eqnarray}

type-$ \beta $ (note the notation:   $f_{1}, f_{2} \rightarrow g_{1},
g_{2}$,)
\begin{eqnarray}
{d^{2}g_{1}\over dy^{2}}+\left[ B -( a+ B
y)^{2}+\lambda^{2}\right]g_{1}=0\,, \nonumber \\
  g_{2} = {1 \over
k +p} \left ( {d \over dy} + (a+ B y) \right ) g_{1} \; .
\nonumber
\label{2.11b}
\end{eqnarray}

\noindent
Because $f_{1}$ and $ g_{1}$ obey the same differential equation, one can set them equal to each other:
\begin{eqnarray}
(\alpha, \beta) \qquad f_{1}(y)  = g_{1} (y) = f (y)  ,
\nonumber
\\
(\alpha) \; f_{2} = {1 \over  k -p} \left ( {d \over dy} + (a+
B y) \right ) f \equiv  {1 \over k -p} g    , \nonumber
\\
( \beta ) \; g_{2} = {1 \over  k +p} \left ( {d \over dy} +
(a+ B y)\right ) f \equiv  {1 \over  k + p}  g .
\nonumber
\label{2.12}
\end{eqnarray}

\noindent Correspondingly, (\ref{2.10}) will read
\begin{eqnarray}
 \Psi_{\epsilon a k, \alpha} =e^{-i\epsilon t}
e^{ia x} e^{ikz} \left | \begin{array}{r}
f (y) \\
(k-p)^{-1}\; g(y)  \\
 \alpha \;
  f (y) \\
\alpha (k-p)^{-1} g(y)
\end{array} \right |,
\nonumber
\\
 \Psi_{\epsilon a k, \beta} =e^{-i\epsilon t} e^{ia
x} e^{ikz} \left | \begin{array}{r}
f (y) \\
 (k+p)^{-1} g (y) \\
\beta\;  f (y) \\
 \beta (k+p)^{-1}   g (y)
\end{array} \right |.
\label{2.13}
\end{eqnarray}

\section{Vanishing the current on the boundaries of the domain between two planes}

Now let us derive condition for vanishing the current $J^{z}$ on the boundaries
$z=-a, z=+a$ of the domain between two planes. We us the above solutions
of the types $(\alpha, \beta)$  and similar to them with the change  $k
\Longrightarrow -k$ (total for all four solutions factor
$e^{-i\epsilon t} e^{ia x}$ will be omitted)
\begin{eqnarray}
\Psi_{1}= \Psi_{\epsilon, a, k, \alpha} =
e^{ikz} \left | \begin{array}{r}
f  \\
(k-p)^{-1}\; g  \\
 \alpha \;
  f  \\
\alpha (k-p)^{-1} g
\end{array} \right |,
\nonumber\\
\Psi_{2}= \Psi_{\epsilon, a, k, \beta} =
e^{ikz} \left | \begin{array}{r}
f (y) \\
 (k+p)^{-1} g  \\
\beta\;  f  \\
 \beta (k+p)^{-1}   g
\end{array} \right |,
\nonumber
\\
\Psi_{3}= \Psi_{\epsilon, a,- k, \alpha} =
e^{-ikz} \left | \begin{array}{r}
f  \\
-(k+p)^{-1}\; g  \\
 \alpha \;
  f  \\
-\alpha (k+p)^{-1} g
\end{array} \right |,\nonumber
\\
\Psi_{4}= \Psi_{\epsilon, a,- k, \beta} = e^{-ikz} \left |
\begin{array}{r}
f  \\
 -(k-p)^{-1} g  \\
\beta\;  f  \\
- \beta (k-p)^{-1}   g
\end{array} \right |.
\label{3.1}
\end{eqnarray}

\noindent Let us make a linear combination of   these:
\begin{eqnarray}
\Phi = A_{1} \Psi_{1} + A_{2} \Psi_{2} + A_{3} \Psi_{3} + A_{4}
\Psi_{4} \; ;
\nonumber
\end{eqnarray}

\noindent the constituents of such a wave function are
\begin{eqnarray}
{\Phi_{1}\over f}  =    e^{ikz}  (A_{1}
+A_{2} ) + e^{-ikz} (A_{3} +A_{4} )        ,\qquad
{\Phi_{3} \over f} =   e^{ikz}  (A_{1}\alpha +A_{2} \beta ) + e^{-ikz}
(A_{3}\alpha  +A_{4} \beta )     ,
\nonumber
\\
{\Phi_{2}\over g}  =  e^{ikz}  ({A_{1}
\over k -p}  +{A_{2} \over k+p} ) - e^{-ikz}( {A_{3} \over k+p } +
{A_{4} \over k-p })       ,
{\Phi_{4}  \over g} =  e^{ikz}  ({A_{1}\alpha \over k -p}  +{A_{2}
\beta \over k+p} )  - e^{-ikz}( {A_{3} \alpha \over k+p } + {A_{4}
\beta \over k-p })      ;
\nonumber
\label{3.2}
\end{eqnarray}

\noindent
note the structure of the current (in the spinor basis)
\begin{eqnarray}
J^{z} = \Phi^{+} \gamma^{0} \gamma^{3} \Phi
= ( \Phi_{1}^{*}
\Phi_{1} -  \Phi_{3}^ {*} \Phi_{3} ) - ( \Phi_{2}^{*} \Phi_{2} -
\Phi_{4}^{*} \Phi_{4})\; . \label{3.3}
\end{eqnarray}

\noindent
The pair   $(\Phi_{1}, \Phi_{3})$ contains one the same factor $f(y)$;
in turn,  the pair   $(\Phi_{2}, \Phi_{4})$ contains other factor $g(y)$.
The current on the boundaries  $z=-a, z =+a$ will vanish (with no
 influence of $y$-dependence), if the following restrictions are
satisfied
$$
\underline{z = -a}\;, \; \Phi_{3} = e^{i\rho } \Phi_{1},
\;  \Phi_{4}
= e^{i\sigma} \Phi_{2} \qquad \Longrightarrow
$$
\begin{eqnarray}
e^{-ika}  (A_{1}\alpha  +A_{2} \beta ) + e^{ika}
(A_{3}\alpha  +A_{4} \beta )
=
 e^{i\rho } [ e^{-ika}  (A_{1} +A_{2} ) + e^{ika} (A_{3} +A_{4} )   ],
\nonumber
\\
 \;[e^{-ika}  ({A_{1}\alpha  \over k -p}  +{A_{2} \beta \over
k+p} )  - e^{ika}( {A_{3} \alpha \over k+p } + {A_{4}  \beta \over
k-p })  ]
=  e^{i\sigma} [ e^{-ika}  ({A_{1} \over k -p}
+{A_{2} \over k+p} ) - e^{ika}( {A_{3} \over k+p } + {A_{4} \over
k-p })  ]  ;
\nonumber
\end{eqnarray}
$$
\underline{z = +a}\;, \; \Phi_{3} = e^{i\mu } \Phi_{1}, \qquad
\Phi_{4} = e^{i\nu} \Phi_{2} \; \Longrightarrow
$$
\begin{eqnarray}
e^{ika}  (A_{1}\alpha  +A_{2} \beta ) + e^{-ika}
(A_{3}\alpha  +A_{4} \beta )  =
 e^{i\mu }  [ e^{ika}  (A_{1} +A_{2} ) + e^{-ika} (A_{3} +A_{4} )   ],
\nonumber
\\
\; [e^{ika}  ({A_{1}\alpha  \over k -p}  +{A_{2} \beta \over
k+p} )  - e^{-ika}( {A_{3} \alpha \over k+p } + {A_{4}  \beta
\over k-p })  ]
= e^{i\nu}  [ e^{ika}  ({A_{1} \over k
-p}  +{A_{2} \over k+p} ) - e^{-ika}( {A_{3} \over k+p } + {A_{4}
\over k-p })  ].
\nonumber
\end{eqnarray}

\noindent
These can be rewritten as  homogeneous linear system with respect
to  $A_{1}, A_{2}, A_{3}, A_{4}$
(let  $K=e^{2iak}$)
\begin{eqnarray}
A_{1} (\alpha - e^{i\rho})   +A_{2} (\beta  - e^{i\rho})   +  A_{3} (\alpha - e^{i\rho})
K  +A_{4} (\beta  - e^{i\rho}) K = 0 \; ,
\nonumber
\\
A_{1}K (\alpha -e^{i\mu} )   +A_{2} K  (\beta-e^{i\mu})  +  A_{3}(\alpha -e^{i\mu}) +A_{4}
( \beta  -e^{i\mu}) = 0 \; ,
\nonumber
\\
  A_{1} (\alpha -e^{i\sigma})  (k +p)  + A_{2} (\beta -e^{i\sigma}) (k-p)  -
 A_{3} K (\alpha-e^{i\sigma}) (k-p )- A_{4} K (\beta -e^{i\sigma})( k+p) =0\; ,
\nonumber
\\
  A_{1} K (\alpha -e^{i\nu}) (k +p)  +A_{2} K  (\beta -e^{i\nu})(k-p )  -
 A_{3} (\alpha -e^{i\nu}) (k-p) - A_{4} ( \beta -e^{i\nu}) (k+p ) = 0\;.
\nonumber
\\
\label{3.6}
\end{eqnarray}

Let us write down explicit form of the matrix for the system
(\ref{3.6})
\begin{eqnarray}
 S=\left | \begin{array}{rrrr}
 (\alpha - e^{i\rho})  &  (\beta  - e^{i\rho})   & (\alpha - e^{i\rho}) K  &  (\beta  - e^{i\rho}) K \\
 (\alpha -e^{i\mu} )K  &    (\beta-e^{i\mu}) K & (\alpha -e^{i\mu}) & ( \beta  -e^{i\mu})  \\
(\alpha -e^{i\sigma})  (k +p)  & (\beta -e^{i\sigma}) (k-p)  & -
   (\alpha-e^{i\sigma}) (k-p )K & -  (\beta -e^{i\sigma})( k+p)K \\
    (\alpha -e^{i\nu}) (k +p)K  &   (\beta -e^{i\nu})(k-p )K &  -
  (\alpha -e^{i\nu}) (k-p)  & - ( \beta -e^{i\nu}) (k+p )
  \end{array} \right |.
 \label{3.7}
 \end{eqnarray}

\section{ Variant with one independent phase}

There exist  6 variants with one independent phase:
\begin{eqnarray}
 \left.  \begin{array}{lrrrrrrrr}
\rho =   & \; \Delta & \; -\Delta & \;  \Delta & \; \Delta & \; \Delta & \;\Delta  &  \;\Delta     &  \;\Delta        \\
\mu =    & \; \Delta & \;  \Delta & \; -\Delta & \; \Delta & \; \Delta & \;\Delta  & \;-\Delta  &  \; -\Delta  \\
\sigma = & \; \Delta & \;  \Delta & \;  \Delta & \;-\Delta & \; \Delta & \;-\Delta &  \;\Delta  &  \; -\Delta    \\
\nu =    & \; \Delta & \;  \Delta & \;  \Delta & \; \Delta & \;-\Delta & \; -\Delta& - \;\Delta &  \;\Delta \\
\end{array} \right. .
\label{4.1}
\end{eqnarray}

\noindent
Let us consider  the first variant in (\ref{4.1}), the main equation takes the form
\begin{eqnarray}
\det S  = \left | \begin{array}{rrrr}
 (\alpha - e^{i\Delta })  &  (\beta  - e^{i\Delta })   & (\alpha - e^{i\Delta }) K  &  (\beta  - e^{i\Delta }) K \\
 (\alpha -e^{i\Delta}  )K  &    (\beta-e^{i\Delta} ) K & (\alpha -e^{i\Delta }) & ( \beta  -e^{i\Delta })  \\
(\alpha -e^{i\Delta })  (k +p)  & (\beta -e^{i\Delta }) (k-p)  & -
   (\alpha-e^{i\Delta }) (k-p )K & -  (\beta -e^{i\Delta })( k+p)K \\
    (\alpha -e^{i\Delta }) (k +p)K  &   (\beta -e^{i\Delta })(k-p )K &  -
  (\alpha -e^{i\Delta }) (k-p)  & - ( \beta -e^{i\Delta }) (k+p )
  \end{array} \right |;
\nonumber
 \label{4.2a}
 \end{eqnarray}

\noindent after elementary transformation it gives
\begin{eqnarray}
\det S = (\alpha - e^{i\Delta})^{2} (\beta  - e^{i\Delta})^{2}
\left | \begin{array}{cccc}
 1  &  1   &  K  &   K \\
 K  &     K & 1 & 1  \\
  (k +p)  & (k-p)  & -     (k-p )K & -  ( k+p)K \\
     (k +p) K &   (k-p )K &  -    (k-p)  & -  (k+p )
  \end{array} \right |.
 \label{4.2b}
 \end{eqnarray}
We see   two evident roots
$
K =e^{2iak} = \pm 1\; .
$
For instance, let  $K=+1$, then
\begin{eqnarray}
\det S = (\alpha - e^{i\Delta})^{2} (\beta  - e^{i\Delta})^{2}
\left | \begin{array}{cccc}
 1  &  1   &  1  &   1 \\
 1  &     1 & 1 & 1  \\
  (k +p)  & (k-p)  & -
    (k-p ) & -  ( k+p) \\
     (k +p)  &   (k-p ) &  -
   (k-p)  & -  (k+p )
  \end{array} \right |.
 \label{4.3}
 \end{eqnarray}

\noindent
 Note that the rank of the matrix in (\ref{4.3}) equals  2.
When  $K=+1$, the system  (\ref{3.6})  reads
\begin{eqnarray}
A_{1} (\alpha - e^{i\Delta})   +A_{2} (\beta  - e^{i\Delta})   +
A_{3} (\alpha - e^{i\Delta})
  +A_{4} (\beta  - e^{i\Delta})  = 0 \; ,
\nonumber
\\
A_{1} (\alpha -e^{i\Delta} )   +A_{2}   (\beta-e^{i\Delta})  +
A_{3}(\alpha -e^{i\Delta}) +A_{4} ( \beta  -e^{i\Delta} ) = 0 \; ,
\nonumber
\\
  A_{1} (\alpha -e^{i\Delta})  (k +p)  + A_{2} (\beta -e^{i\Delta}) (k-p)  -
 A_{3}  (\alpha-e^{i\Delta}) (k-p )- A_{4}  (\beta -e^{i\Delta})( k+p) =0\; ,
\nonumber
\\
  A_{1}  (\alpha -e^{i\Delta}) (k +p)  +A_{2}   (\beta -e^{i\Delta})(k-p )  -
 A_{3} (\alpha -e^{i\Delta}) (k-p) - A_{4} ( \beta -e^{i\Delta}) (k+p ) = 0\;.
\nonumber
\end{eqnarray}

\noindent
In accordance  with  the property $\mbox{rank}\; S = 2$, we get only two different equations
\begin{eqnarray}
A_{1} (\alpha -e^{i\Delta} )   +A_{2}   (\beta-e^{i\Delta})  +
A_{3}(\alpha -e^{i\Delta}) +A_{4} ( \beta  -e^{i\Delta} ) = 0 \; ,
\nonumber
\\
  A_{1} (\alpha -e^{i\Delta})  (k +p)  + A_{2} (\beta -e^{i\Delta}) (k-p)  -
 A_{3}  (\alpha-e^{i\Delta}) (k-p )- A_{4}  (\beta -e^{i\Delta})( k+p) =0\; .
  \label{4.4}
 \end{eqnarray}

\noindent
 They read as a linear system with respect to $A_{3}, A_{4}$;
  its solution is
\begin{eqnarray}
A_{3} = - {k+p \over p }A_{1}
 - {k \over p} { (\beta- e^{i\Delta} )\over  (\alpha - e^{i\Delta})}\; A_{2} \; ,
\qquad A_{4} =  {k \over p} {(\alpha - e^{i\Delta}) \over  (\beta-
e^{i\Delta})} \; A_{1} + {k-p \over p } A_{2} \; . \label{4.4c}
\end{eqnarray}

\noindent
The corresponding wave function and quantization rule for
 $k$ are
 \begin{eqnarray}
J^{z} (z = \pm a)=0, \qquad K = e^{2ika} = +1, \qquad k = {\pi
\over a }  n,\; n = 0, \pm 1, \pm 2, ... ;\qquad\qquad
\nonumber
\\
 \Phi = A_{1} \Psi_{1} + A_{2} \Psi_{2} +
 \left ( - {(k+p) \over p }A_{1}
 - {k \over p} { (\beta- e^{i\Delta} )\over  (\alpha - e^{i\Delta})}\; A_{2} \right ) \Psi_{3} +
 \left (  {k \over p} {(\alpha - e^{i\Delta}) \over  (\beta- e^{i\Delta})} \; A_{1} +
{k-p \over p } A_{2} \right ) \Psi_{4}\;. \label{4.5a}
\end{eqnarray}

\noindent It should be emphasized that the phase  $\Delta$  does not determine quantization
of the $k$, instead it only influences on the coefficients of the  linear combination
in (\ref{4.5a}). A simplest choice for  the phase is $\Delta = 0$:
\begin{eqnarray}
\Phi = A_{1} \Psi_{1} + A_{2} \Psi_{2} + \left ( - {(k+p) \over p
}A_{1}
 + \alpha^{-1} {k \over p} \; A_{2} \right ) \Psi_{3} +
  \left (  - \alpha {k \over p}  \; A_{1} +
{k-p \over p } A_{2} \right ) \Psi_{4}\;.
\label{4.5b}
\end{eqnarray}

To examine  two remaining solutions, let us turn back to the  initial polynomial (\ref{4.2b})
and find all its roots.  Explicitly, equation  $\det S = 0$ reads
\begin{eqnarray}
 ( K^{2}-1 )^{2}   =0 \qquad \Longrightarrow \qquad
K=+1, -1,+1, -1\; .
  \label{4.6b}
 \end{eqnarray}

Let us consider the second variant in  (\ref{4.1}):
\begin{eqnarray}
 S=\left | \begin{array}{cccc}
 (\alpha - e^{-i\Delta})  &  (\beta  - e^{-i\Delta})   & (\alpha - e^{-i\Delta}) K  &  (\beta  - e^{-i\Delta} ) K \\
 (\alpha -e^{i\Delta} )K  &    (\beta-e^{i\Delta}) K & (\alpha -e^{i\Delta}) & ( \beta  -e^{i\Delta})  \\
(\alpha -e^{i\Delta})  (k +p)  & (\beta -e^{i\Delta}) (k-p)  & -
   (\alpha-e^{i\Delta}) (k-p )K & -  (\beta -e^{i\Delta})( k+p)K \\
    (\alpha -e^{i\Delta}) (k +p)K  &   (\beta -e^{i\Delta})(k-p )K &  -
  (\alpha -e^{i\Delta}) (k-p)  & - ( \beta -e^{i\Delta}) (k+p )
  \end{array} \right |,
\nonumber
\end{eqnarray}

\noindent or
  \begin{eqnarray}
 \det S =(\alpha -e^{i\Delta}) ^{2} (\beta -e^{i\Delta} )^{2}
\left | \begin{array}{cccc}
 {\alpha - e^{-i\Delta} \over
  \alpha - e^{i\Delta}  }   &  { \beta  - e^{-i\Delta} \over \beta  - e^{i\Delta}}   &
 {\alpha - e^{-i\Delta} \over \alpha - e^{i\Delta} }K  &
  {\beta  - e^{-i\Delta}  \over \beta  - e^{i\Delta} }K \\
 K  &     K & 1 & 1  \\
  (k +p)  &  (k-p)  & -
    (k-p )K & -  ( k+p)K \\
     (k +p)K  &   (k-p )K &  -
   (k-p)  & - (k+p )
  \end{array} \right |.
\label{4.7a}
\end{eqnarray}

\noindent In  (\ref{4.7a}), the rows  3  and 4  are proportional if $K=+1, K=-1$, which means that
the values $K= +1, -1$ are the roots of the polynomial. At this, the rank of the matrix equal to 3,
because the left upper block  $ S_{3
\times 3}$  has a non-zero determinant:
\begin{eqnarray}
K= +1, \qquad \det S_{3 \times 3} = {2k   (1 -\alpha^{2})
  \over x ( \alpha\,x-1 )  ( x -\alpha ) } (1-x^{2}),  \qquad e^{i\Delta} = x \; ;
\nonumber
\\
K= -1, \qquad \det S_{3 \times 3} = {2k  (1 -\alpha^{2})  \over x
( \alpha\,x-1 )  ( x -\alpha ) }(1-x^{2}),
 \qquad e^{i\Delta} = x\; .
\label{4.8}
\end{eqnarray}

\noindent
 However, this determinant vanishes if $x= +1, -1 \;
  (\Delta =0, \pi )$ , and in this case the rank of the matrix equals to 2.

 Let us detail in the case  $\det a_{3 \times 3} \neq 0$
the linear when $K=+1$;  the fourth equation coincides with the third and we have only thee different
equations
\begin{eqnarray}
A_{1} (\alpha - e^{-i\Delta})   +A_{2} (\beta  - e^{-i\Delta})   +
A_{3} (\alpha - e^{-i\Delta})
  =- A_{4} (\beta  - e^{-i\Delta})  \; ,
\nonumber
\\
A_{1} (\alpha -e^{i\Delta} )   +A_{2}   (\beta-e^{i\Delta})  +
A_{3}(\alpha -e^{i\Delta}) =- A_{4} ( \beta  -e^{i\Delta})  \; ,
\label{4.9a}
\\
  A_{1} (\alpha -e^{i\Delta})  (k +p)  + A_{2} (\beta -e^{i\Delta}) (k-p)  -
 A_{3}  (\alpha-e^{i\Delta}) (k-p ) =  A_{4}  (\beta -e^{i\Delta})( k+p) \; .
\nonumber
\end{eqnarray}

\noindent With the use of notation  $e^{i\Delta} = x$, solution of
these equation looks (as should be expected, the formulas are
meaningless when   $x=\pm 1$.)
\begin{eqnarray}
A_{1} =  {  2k{x}^{3} \alpha+ (
-k{\alpha}^{2}+p{\alpha}^{2}-3\,k+p ) {x}^{2}+
 2\alpha (  k-2 p ) x-k{\alpha}^{2}+p{\alpha}^{2
}+k+p  \over 2\alpha ( x-1 )  ( x+1 )
 ( x-\alpha  k)}   \; A_{4} =c_{1} A_{4},
\nonumber
\\
A_{2} = -{\frac {
 \left( x-\alpha \right) x}{ \left( x-1 \right)  \left( x+1 \right) }} \;A_{4} = c_{2} A_{4},\hspace{40mm}
\label{4.9b}
\\
A_{3} = -{  2k{x}^{3}\alpha+
 ( -k{\alpha}^{2}+p{\alpha}^{2}-k+p ) {x}^{2} -2\alpha + ( k +2 p ) x + k{\alpha}^{2}+p{\alpha}^{2}+k+p
 \over 2 \alpha ( x-1 )  ( x+1 )  ( x-
\alpha  k)}  \; A_{4} = c_{3} A_{4} \; .
\nonumber
\end{eqnarray}

\noindent
The corresponding wave function and quantization rule for $k$ are
\begin{eqnarray}
\Psi = A_{4} \left ( c_{1}\Psi_{1} + c_{2}\Psi_{2}+ c_{3}\Psi_{3}
+ \Psi_{4} \right ) ; \label{4.9c}
\end{eqnarray}

\noindent at this an arbitrary phase parameter  $x = e^{i\Delta}$  influences
 $c_{1}, c_{2}, c_{3}$ in (\ref{4.9c}).

Turning back to general case (\ref{4.7a}), we see that  $\det S =0$ is  a bi-quadratic equation
(let   $\Lambda = K^{2}$)
\begin{eqnarray}
\left(
k{x}^{2}{\alpha}^{2}+p{x}^{2}{\alpha}^{2}-k{x}^{2}-k{\alpha}^{
2}+p{x}^{2}-4\,px\alpha+p{\alpha}^{2}+k+p \right) {\Lambda}^{2}
\nonumber
\\
+\left( -2\,
p{x}^{2}{\alpha}^{2}-2\,p{x}^{2}+8\,px\alpha-2\,p{\alpha}^{2}-2\,p
 \right) \Lambda
 \nonumber
 \\
 -
 k{x}^{2}{\alpha}^{2}+p{x}^{2}{\alpha}^{2}+k{x}^{2}+k{\alpha
}^{2}+p{x}^{2}-4\,px\alpha+p{\alpha}^{2}-k+p=0\; ; \label{4.10a}
\end{eqnarray}

\noindent its roots are
\begin{eqnarray}
\Lambda_{1} =1, \qquad \Lambda_{2} = -{\frac {  [   ( {x}^{2}-1
) {\alpha}^{2}-{x}^{2}+1
 ]  k+  [ ( -{x}^{2}-1 ) {\alpha}^{2}+4\,\alpha\,
x-{x}^{2}-1 ] ) p}{ [  ( {x}^{2}-1 )
{\alpha}^{2}- {x}^{2}+1  ] k+  [  ( {x}^{2}+1
) {\alpha}^{2}-4\, \alpha\,x+{x}^{2}+1 ] p}}\;.
\label{4.10b}
\end{eqnarray}

In order to obtain a complex conjugate  $\Lambda_{2} ^{*}(x)$, it is enough in  $\Lambda_{2}(x)$ to make  the
formal change $x \Longrightarrow x^{-1}$, after that by direct calculation one proves
identity
$\Lambda_{2} \Lambda^{*}_{2}  = \Lambda_{2}(x) \Lambda_{2}(x^{-1}) = +1$.
 This means that the root  $\Lambda_{2}$ is of a  phase type and it is appropriate to
give quantization rule for $k$:
\begin{eqnarray}
e^{4iak} = \Lambda_{2} (\epsilon, p, k ,e^{i\Delta} ) \; . \label{4.11b}
\end{eqnarray}

\noindent It should be specially noted that eq.  (\ref{4.11b})
contain an arbitrary phase  $x= e^{i\Delta}$, and it can be solved
only numerically, which makes it of  little significance in the
context of theoretical analysis. However if $x=+1, -1 (\Delta= 0,
\pi)$, the root $\Lambda_{2} =1$.

Consider the variant 3 in  (\ref{4.1}):
\begin{eqnarray}
\det \left| \begin {array}{cccc} X   &   1&  KX  &   K
\\  -K  &   -KX &    -1 &   -X
\\  X \left( k+p \right) &  k-p &   -KX \left( k-p \right)  &   - \left( k+p \right) K \\
X \left( k+p \right) K  &   K \left( k-p \right) & -X \left( k-p
\right) & -k-p
\end {array} \right| =0 \; ,\qquad  x= e^{i\Delta} , \; {\alpha - x \over 1 - \alpha x } = X;
\label{4.12a}
\end{eqnarray}

\noindent
 it reduces to a  bi-quadratic equation
(let  $ K^{2}=\Lambda$)
\begin{eqnarray}
  [   ( k + p  ) X^2 - k + p  ] \Lambda^2
- 2p  ( X^2 + 1  ) \Lambda - [   ( k-p  ) {X}^{2}- k - p  ]   =0 \; ;
\label{4.12b}
\end{eqnarray}

\noindent its roots are
\begin{eqnarray}
\Lambda_{1}=1\; , \qquad \Lambda_{2}= { \frac { ( - k + p  ) {X}^{2}+k+p
      }
      {
( k+p  ) {X}^{2}-k+p       }} \; , \qquad
\Lambda_{2}\; \Lambda_{2}^*= \Lambda_{2}(X) \; \Lambda_{2}({1 \over X}) = 1 \; .
\label{4.12c}
\end{eqnarray}

Consider  the  variant 4 in  (\ref{4.1}):
\begin{eqnarray}
\det \left|
\begin {array}{cccc}
X   &   1   &   KX  &   K\\
  KX    &   K   &   X   &   1
\\ -k-p &   -X  ( k-p  ) &  K ( k-p )   &   X ( k+p ) K
\\  X ( k+p ) K &   K ( k-p ) & -X ( k-p ) & -k-p
\end {array}
\right| =0 \; ,\qquad
x= e^{i\Delta} , \; {\alpha - x \over 1 - \alpha x } = X \;. \label{4.13a}
\end{eqnarray}

\noindent Further we get a bi-quadratic equation (let  $K^{2} = \Lambda$)
\begin{eqnarray}
[  ( k+p ) {X}^{2} - k+p ]\Lambda^2 - 2p ( X^2 + 1 ) \Lambda - [( k-p )X^2-k-p
]=0 \; ; \label{4.13b}
\end{eqnarray}

\noindent its roots  are
\begin{eqnarray}
\Lambda_{1}=1, \qquad \Lambda_{2}= {\frac { (   -k + p  ){X}^{2} + k + p } { (
k + p ) {X}^{2} - k + p } }\; ; \qquad
\Lambda_{2}\; \Lambda_{2}^* = \Lambda_{2} (X) \; \Lambda_{2} ({1\over X}) = 1 \; .
\label{4.13c}
\end{eqnarray}

Consider the  variant 5 in (\ref{4.1}):
\begin{eqnarray}
\det \left|
\begin {array}{cccc}
X   &   1   &   K   X   &   K
\\ KX   &   K   &   X   &   1
\\  X  ( k+p  ) &   k-p &   -KX  ( k-p  )   &   -  ( k+p  ) K
\\ - \left( k+p \right) K   &- ( k-p ) K & k-p  &   k+p
\end {array}
\right| =0\; ; \label{4.14a}
\end{eqnarray}

\noindent we arrive at a  bi-quadratic equation
with respect to $\Lambda = K^{2}$
$$
-   [  ( k-p ) X  -   k-p ]{\Lambda}^{2} -   2p ( X-1 ) \Lambda +   [  ( k+p )
X-k+p ]  =0 \; ; \eqno(4.14b)
$$

\noindent its roots are
\begin{eqnarray}
\Lambda_{1}=1, \qquad \Lambda_{2}= {\frac { -X( k+p ) + k-p} {  X( k-p ) -
k-p} } \; , \qquad \Lambda_{2}\; \Lambda_{2}^*=1 \; . \label{4.14c}
\end{eqnarray}

Consider the  variant 6 in (\ref{4.1}):
\begin{eqnarray}
\det \left|
\begin {array}{cccc}
X  &    1   &   KX  &   K   \\
KX  &   K   &   X  & 1
\\ - k-p &   -X ( k-p ) & ( k-p ) K  &   X ( k+p ) K
\\ - ( k+p ) K  &   - ( k-p ) K &   k  -    p   &   k+p
\end {array}
\right| = 0, \qquad
x= e^{i\Delta} , \qquad {\alpha - x \over 1 - \alpha x } = X \; . \label{4.15a}
\end{eqnarray}

\noindent Further we get a bi-quadratic equation (let $K^{2} = \Lambda$)
\begin{eqnarray}
[  ( k+p ) {X}^{2}-k+p ]
 [  ( k-p ) X-k-p ]
 \Lambda^{2}\hspace{30mm}
\nonumber
\\
-2[  ( {k}^{2}-{p}^{2} )
 ( {X}^{2}+1 ) -2X{k}^{2} ]
  ( X+1 ) \Lambda+
 [  ( k-p ) {X}^{2}-k-p ]
 [  ( k+p ) X-k+p ] =0 \; ,
\label{4.15b}
\end{eqnarray}

\noindent with the roots
\begin{eqnarray}
\Lambda_{1}=1, \qquad \Lambda_{2}= {\frac {  [ X ( k+p ) -k+p ]
 [  ( k-p ) {X}^{2}-k-p ] }
{ [  ( k+p ) {X}^{2}-k+p ]
  [ X ( k-p ) -k-p ]
} }\; , \qquad \Lambda_{2}\; \Lambda_{2}^*=1 \; .
\label{4.15c}
\end{eqnarray}

Consider the  variant 7 in (\ref{4.1})
\begin{eqnarray}
\det S =\left | \begin{array}{rrrr}
 (\alpha - e^{i\Delta})  &  (\beta  - e^{i\Delta})   & K(\alpha - e^{i\Delta})   &  K(\beta  - e^{i\Delta})  \\
 K(\alpha -e^{-i\Delta} )  &    K(\beta-e^{-i\Delta})  & (\alpha -e^{-i\Delta}) & ( \beta  -e^{-i\Delta})  \\
(\alpha -e^{i\Delta})  (k +p)  & (\beta -e^{i\Delta}) (k-p)  & -
   K(k-p )(\alpha-e^{i\Delta})  & -  K( k+p)(\beta -e^{i\Delta}) \\
   K (k +p) (\alpha -e^{-i\Delta})   &   K (k-p )(\beta -e^{-i\Delta}) &  -
  (k-p) (\alpha -e^{-i\Delta})   & - (k+p )( \beta -e^{-i\Delta})
  \end{array} \right |
\label{4.16a}
\end{eqnarray}

\noindent or with notation $e^{i\Delta} = x$:
\begin{eqnarray}
 {1 \over x^{2}\alpha^{2} } \left |
\begin{array}{rrrr}
 (\alpha - x)  &  (1  - \alpha x)   & K (\alpha - x)   &  K (1 - \alpha x)   \\
 -K(1 -\alpha x)  &    -K(\alpha - x)
  & -(1 -\alpha x)   & -(\alpha - x)    \\
(k +p) (\alpha -x )    & (k-p)   (1 - \alpha x)    & -
   K(k-p)   (\alpha-x)  & - K( k+p) (1 - \alpha x )  \\
    -K (k +p) (1 -\alpha  x )  &  -K (k-p ) (\alpha - x)   &
  (k-p) (1 -\alpha  x )    &  (k+p ) (\alpha - x)
  \end{array} \right |=0\; .
\nonumber
\end{eqnarray}

\noindent Further with the use of notation
$
{\alpha - x \over 1 - \alpha x } = X \; ,
$
 we arrive at
 \begin{eqnarray}
 {(1 - \alpha x )^{4} \over x^{2}\alpha^{2} } \left |
\begin{array}{rrrr}
 X  &  1  & K X   &  K   \\
 -K  &    -K X
  & -1   & - X    \\
X(k +p)     & (k-p)      & -
   K X (k-p)     & - K( k+p)   \\
    -K (k +p)   &  -K X(k-p )    &
  (k-p)    &  X(k+p )
  \end{array} \right |=0\; .
\label{4.16d}
\end{eqnarray}

\noindent So we get a bi-quadratic equation (let it be $K^{2} = \Lambda $)
\begin{eqnarray}
\Lambda^{2} - \Lambda [ L (W-2)+2] +1 = 0\;, \qquad
L=k^{2}/p^{2}, \qquad W = X^{2} + {1 \over X^{2}} \; .
\label{4.17a}
\end{eqnarray}

\noindent The roots are
\begin{eqnarray}
\Lambda_{1} =1 + {L(W-2) \over 2} + {\sqrt{L(W-2) [ 4 + L(W-2)]} \over 2}\; ,
\nonumber
\\
\Lambda_{2} =1 + {L(W-2) \over 2} - {\sqrt{L(W-2) [ 4 + L(W-2)]} \over 2}\; .
\label{4.17b}
\end{eqnarray}

\noindent
Due to easily checked identities
$
W^{*} = W
$
and
\begin{eqnarray}
\Lambda_{1}^{*} =\Lambda_{1} =1 + {L(W-2) \over 2} + {\sqrt{L(W-2) [ 4 + L(W-2)]} \over 2}\; ,
\nonumber
\\
\Lambda_{2}^{*} = \Lambda_{2} =1 + {L(W-2) \over 2} - {\sqrt{L(W-2) [ 4 + L(W-2)]} \over 2}\; .
\nonumber
\end{eqnarray}

\noindent
we derive identities
$
\Lambda_{1}^{*} \Lambda_{1}=1  \; , \; \Lambda_{2}^{*} \Lambda_{2}=1\; .
$
So, all the roots are complex number of the unit length.

Consider  variant 8 in (\ref{4.1}):
\begin{eqnarray}
\det S = \left | \begin{array}{rrrr}
 (\alpha - e^{i\Delta })  &  (\beta  - e^{i\Delta })   & (\alpha - e^{i\Delta }) K  &  (\beta  - e^{i\Delta }) K \\
 (\alpha -e^{-i\Delta}  )K  &    (\beta-e^{-i\Delta} ) K & (\alpha -e^{-i\Delta }) & ( \beta  -e^{-i\Delta })  \\
(\alpha -e^{-i\Delta })  (k +p)  & (\beta -e^{-i\Delta }) (k-p)  &
-
   (\alpha-e^{-i\Delta }) (k-p )K & -  (\beta -e^{-i\Delta })( k+p)K \\
    (\alpha -e^{i\Delta }) (k +p)K  &   (\beta -e^{i\Delta })(k-p )K &  -
  (\alpha -e^{i\Delta }) (k-p)  & - ( \beta -e^{i\Delta }) (k+p )
  \end{array} \right |;
 \nonumber
 \label{4.21a}
 \end{eqnarray}

\noindent with notation
$
e^{i\Delta }=x \; , \;  {\alpha - x \over 1  -\alpha x } = X
$
we rewrite this  as
\begin{eqnarray}
 \left | \begin{array}{rrrr}
X &  \alpha^{-1}   & X K  &  \alpha^{-1} K \\
 -x^{-1}K  &   - X\alpha^{-1}x^{-1}K & -x^{-1}  & -X \alpha^{-1}x^{-1}  \\
 -{x}^{-1}  (k +p)  & -X\alpha^{-1}x^{-1}  (k-p)  & x^{-1} (k-p )K &  X( k+p)\alpha^{-1}x^{-1}K   \\
X(k +p)K  & \alpha^{-1}(k-p )K &  -  X(k-p)  & - \alpha^{-1} (k+p
)
  \end{array} \right |=0 \; ;
\nonumber
\end{eqnarray}

\noindent and further
as
\begin{eqnarray}
\det S  = {{(1 -\alpha x)^4} \over x^2\alpha^2} \left |
\begin{array}{rrrr}
X &    1  & X K  &   K \\
 - K  &   - X K & - 1  & -X   \\
 - (k +p)  & -X  (k-p)  & (k-p )K &  X( k+p) K   \\
X(k +p)K  &  (k-p )K &  -  X(k-p)  & -  (k+p )
  \end{array} \right | =0 \; .
 \label{4.21c}
 \end{eqnarray}

\noindent Which results in  a bi-quadratic equation
\begin{eqnarray}
 \left [   (  {X}^{2}  ( k+p )  -   k      +    p ) {K}^{2}   +   {X}^{2} ( k-p ) -k-p  \right ] ^{2}=0\; ;
\label{4.22a}
\end{eqnarray}

\noindent its 2-multiple  roots are  given by (let $K^{2}=\Lambda$)
\begin{eqnarray}
\Lambda={\frac {  ( -k+p  ) {X}^{2}    +k + p} {( k+p  ) {X}^{2}
-k + p}
  } \;, \qquad \Lambda \; \Lambda^*=1 \; .
\label{4.22b}
\end{eqnarray}

\section{Cases with two independent phases  }

There exist only 4 substantially different possibilities
to fix four phases  with the use of 2 parameters:
\begin{eqnarray}
 \left. \begin{array}{lrrrr}
\rho =   & \; \Delta & \; \Delta & \;  \Delta & \; \Delta        \\
\mu =    & \; \Delta & \;  \Delta & \; -\Delta & \; -\Delta   \\
\sigma = & \; W & \;  W & \;  W & \;W \\
\nu =    & \; W & \;  -W & \;  W & \; -W
\end{array} \right.  .
\label{5.1}
\end{eqnarray}

\noindent
Let us  the  variant 1 in  (\ref{5.1}):
\begin{eqnarray}
 \det S = \left | \begin{array}{rrrr}
 (\alpha - e^{i\Delta})  &  (\beta  - e^{i\Delta})   & (\alpha - e^{i\Delta}) K  &  (\beta  - e^{i\Delta}) K \\
 (\alpha -e^{i\Delta} )K  &    (\beta-e^{i\Delta}) K & (\alpha -e^{i\Delta}) & ( \beta  -e^{i\Delta})  \\
(\alpha -e^{iW})  (k +p)  & (\beta -e^{iW}) (k-p)  & -
   (\alpha-e^{iW}) (k-p )K & -  (\beta -e^{iW})( k+p)K \\
    (\alpha -e^{iW}) (k +p)K  &   (\beta -e^{iW})(k-p )K &  -
  (\alpha -e^{iW}) (k-p)  & - ( \beta -e^{iW}) (k+p )
  \end{array} \right |.
 \nonumber
 \end{eqnarray}

\noindent With notation
$
e^{i\Delta} = x \; ,\;e^{iW}= y \; ,
$ it reads
\begin{eqnarray}
 \det S = \left | \begin{array}{rrrr}
 (\alpha - x)  &  (1  - \alpha x)  \alpha^{-1} & (\alpha - x) K  &  (1  - \alpha x)\alpha^{-1} K \\
 (\alpha -x )K  &    (1 - \alpha x)\alpha^{-1}  K & (\alpha -x) & ( 1   - \alpha x) \alpha^{-1} \\
(\alpha -y)  (k +p)  & (1  -\alpha y)\alpha^{-1} (k-p)  & -
   (\alpha-y) (k-p )K & -  (1  - \alpha y) \alpha^{-1}( k+p)K \\
    (\alpha -y) (k +p)K  &   (1 - \alpha y)\alpha^{-1} (k-p )K &  -
  (\alpha -y) (k-p)  & - ( 1 - \alpha y) \alpha^{-1}(k+p )
  \end{array} \right |;
 \nonumber
 \end{eqnarray}

\noindent introducing notation
$
{\alpha - x \over 1 - \alpha x } = X, \; {\alpha - y \over 1 -
\alpha x } = Y \; ;
$ after simple manipulation we arrive at
\begin{eqnarray}
\det S = {(1  - \alpha x)^{2} (1 - \alpha y)^{2}\over \alpha^{2}}
\left | \begin{array}{rrrr}
 X  &  1  &  X K  &   K \\
  X K  &      K &  X & 1 \\
Y  (k +p)  &  (k-p)  & -     Y (k-p )K & -  ( k+p)K \\
     Y (k +p)K  &    (k-p )K &  -
   Y (k-p)  & - (k+p )
  \end{array} \right |.
 \label{5.2c}
 \end{eqnarray}

\noindent Equation  $\det S=0$ turns to have the form
\begin{eqnarray}
[ X(k+p)-Y(k-p )]\; [ X(k-p)-Y(k+p) ]\; ( K-1) ^{2}( K+1) ^{2}=0
\; ; \label{5.3a}
\end{eqnarray}

\noindent its roots are
$
K=1\; ,\quad K=1 \; , \quad K=-1 \; , \quad K=-1 \; .
$

Let us consider  the variant  2
 in  (\ref{5.1})
$$
\det S
$$
\begin{eqnarray}
=\left | \begin{array}{rrrr}
 (\alpha - e^{i\Delta})  &  (\beta  - e^{i\Delta})   & (\alpha - e^{i\Delta}) K  &  (\beta  - e^{i\Delta}) K \\
 (\alpha -e^{i\Delta} )K  &    (\beta-e^{i\Delta}) K & (\alpha -e^{i\Delta}) & ( \beta  -e^{i\Delta})  \\
(\alpha -e^{iW})  (k +p)  & (\beta -e^{iW}) (k-p)  & -
   (\alpha-e^{iW}) (k-p )K & -  (\beta -e^{iW})( k+p)K \\
    (\alpha -e^{-iW}) (k +p)K  &   (\beta -e^{-iW})(k-p )K &  -
  (\alpha -e^{-iW}) (k-p)  & - ( \beta -e^{-iW}) (k+p )
  \end{array} \right |.
 \label{5.3a}
 \end{eqnarray}

\noindent With notation
 $ e^{i\Delta} = x,\; e^{iW}= y, $
it reads
$$
\det S =
$$
\begin{eqnarray}
\hspace{-5mm}
= \left | \begin{array}{rrrr}
 (\alpha - x)  &  (1 -  \alpha  x) \alpha^{-1}   & (\alpha - x) K  &  (1 - \alpha x) \alpha^{-1}K \\
 (\alpha -x )K  &    (1 -\alpha x)\alpha^{-1}  K & (\alpha -x) & ( 1  -\alpha x) \alpha^{-1} \\
(\alpha -y)  (k +p)  & (1  - \alpha y)\alpha^{-1} (k-p)  & -   ( \alpha-y) (k-p )K & -  (1 - \alpha y)\alpha^{-1} ( k+p)K \\
   - ( 1 - \alpha y ) y^{-1}(k +p)K  &   -(\alpha -y) \alpha^{-1}y^{-1} (k-p )K &
   (k-p) ( 1 - \alpha y ) y^{-1} & (\alpha -y) \alpha^{-1}y^{-1} (k+p )
  \end{array} \right |,
  \nonumber
  \end{eqnarray}

\noindent from whence we arrive at
\begin{eqnarray}
\det \left | \begin{array}{rrrr}
  X         &  1    &  X K  &  K \\
 X K        &  K     &  X & 1 \\
 Y  (k +p)  &  (k-p)  & -     Y (k-p )K & -   ( k+p)K \\
-  (k +p)K  & - Y (k-p )K &    (k-p)  &  Y (k+p )
  \end{array} \right | = 0\; .
  \label{5.3b}
  \end{eqnarray}

\noindent In explicit form, equation  $\det S =0$ looks as
\begin{eqnarray}
-( K^2-1) \left  \{ [  ( k^{2}-p^{2} )Y(  X^{2}   +  1  )   -  ( Y^{2} (
k-p)^{2}   + (k+p)^{2}  ) X    ] K^{2}     \right.
\nonumber
\\
\left. - ( {k}^{2}  -   {p}^{2} ) Y({X}^{2}   +  1)   +   [ Y^{2} (  k +
p)^{2}   +    (k - p)^{2}  ]  X  \right \} =0 \; ; \label{5.4a}
\end{eqnarray}

\noindent its roots are
\begin{eqnarray}
K_{1,2}=\pm1,\quad K_{3,4}=\pm\sqrt {[ YX(k+p)-k+p  ] \;  [
X(k-p)-Y(k+p)] \over {[ YX(k-p)-k-p  ] \;  [ X(k+p)-Y(k-p)]}}\; .
\label{5.4b}
\end{eqnarray}

\noindent We easily  derive identities
$
K_{3} \; K_{3}^{*}=1\; , \quad K_{4} \; K_{4}^{*}=1 \; .
$

Let us consider  the variant  3  in (\ref{5.1}):
\begin{eqnarray}
\det S = \left | \begin{array}{rrrr}
 (\alpha - e^{i\Delta})  &  (\beta  - e^{i\Delta})   & (\alpha - e^{i\Delta}) K  &  (\beta  - e^{i\Delta}) K \\
 (\alpha -e^{-i\Delta} )K  &    (\beta-e^{-i\Delta}) K & (\alpha -e^{-i\Delta}) & ( \beta  -e^{-i\Delta})  \\
(\alpha -e^{iW})  (k +p)  & (\beta -e^{iW}) (k-p)  & -
   (\alpha-e^{iW}) (k-p )K & -  (\beta -e^{iW})( k+p)K \\
    (\alpha -e^{iW}) (k +p)K  &   (\beta -e^{iW})(k-p )K &  -
  (\alpha -e^{iW}) (k-p)  & - ( \beta -e^{iW}) (k+p )
  \end{array} \right |=0;
 \nonumber
 \label{5.5a}
 \end{eqnarray}

\noindent with notation
$ e^{i\Delta} = x,\; e^{iW}= y $ it reads
 \begin{eqnarray}
\det S = \left | \begin{array}{rrrr}
 (\alpha - x)  &  (1  - \alpha x) \alpha^{-1}  & (\alpha - x) K  &  (1  - \alpha x) \alpha^{-1} K \\
 - (1 -\alpha x)x^{-1} K  &    -(\alpha -x)\alpha^{-1} x^{-1}  K & -(1 -\alpha x)x^{-1} &
 -(\alpha -x)\alpha^{-1} x^{-1}  \\
(\alpha -y)  (k +p)  & (1 - \alpha y) \alpha^{-1} (k-p)  & -
   (\alpha-y) (k-p )K & -  (1 - \alpha y) ( k+p)K \\
    (\alpha -y) (k +p)K  &   (1  - \alpha y) \alpha^{-1} (k-p )K &  -
  (\alpha -y) (k-p)  & - ( 1 - \alpha y) \alpha^{-1} (k+p )
  \end{array} \right |=0.
 \nonumber
 \end{eqnarray}

\noindent From this we derive
\begin{eqnarray}
\det \left | \begin{array}{rrrr}
  X   &  1  & \ X  K  &   K \\
 -  K  &    -X  K & -1  & -X  \\
 Y  (k +p)  &  (k-p)  & -     Y (k-p )K & -   ( k+p)K \\
    Y (k +p)K  &   (k-p )K &  -  Y (k-p)  & -  (k+p )
  \end{array} \right | =0 \; .
 \label{5.5b}
 \end{eqnarray}

 \noindent
 In explicit form, equation $\det S =0$ is
 \begin{eqnarray}
  ( K^2-1  ) \{
[ ( X^{2} (k+p)^{2}  +  (k-p)^{2})Y
  -(k^{2}-p^{2}) ({Y}^{2} +1 ) X
 ] {K}^{2} -
\nonumber
\\
-   [(k-p)^{2}{X}^{2}   +   (k+p)^{2} ] Y +  (k^{2}-p^{2})(Y^{2}+1)
X     \} =0 \; ; \label{5.6a}
\end{eqnarray}

\noindent its roots are
$$
K_{1,2}=\pm1,\quad K_{3,4}=\pm\sqrt { [ YX(k-p)-k-p  ] \; [
X(k-p)-Y(k+p)) \over { [ YX(k+p)-k+p  ] \; [  X(k+p)-Y(k-p)] }} \; ;
\eqno(5.6b)
$$
it is readily verified  that
$
K_{3} \; K_{3}^{*}=1 \; , \; K_{4} \; K_{4}^{*}=1 \; .
$

Consider  the variant    4  in (\ref{5.1}):
\begin{eqnarray}
\det S = \left | \begin{array}{rrrr}
 (\alpha - e^{i\Delta})  &  (\beta  - e^{i\Delta})   & (\alpha - e^{i\Delta}) K  &  (\beta  - e^{i\Delta}) K \\
 (\alpha -e^{-i\Delta} )K  &    (\beta-e^{-i\Delta}) K & (\alpha -e^{-i\Delta}) & ( \beta  -e^{-i\Delta})  \\
(\alpha -e^{iW})  (k +p)  & (\beta -e^{iW}) (k-p)  & -
   (\alpha-e^{iW}) (k-p )K & -  (\beta -e^{iW})( k+p)K \\
    (\alpha -e^{-iW}) (k +p)K  &   (\beta -e^{-iW})(k-p )K &  -
  (\alpha -e^{-iW}) (k-p)  & - ( \beta -e^{-iW}) (k+p )
  \end{array} \right |;
\nonumber
\label{5.7a}
\end{eqnarray}

\noindent which can be translated to to form
\begin{eqnarray}
\det
 \left | \begin{array}{rrrr}
X  &  1 &  X K  &   K \\
 -  K  &    - X  K & -1 &
 -  X   \\
 Y  (k +p)  & (k-p)  & -     Y  (k-p )K & -  ( k+p)K \\
   -  (k +p)K  &   - Y (k-p )K &
    (k-p)  & Y (k+p )
  \end{array} \right |=0 \;.
 \label{5.7b}
 \end{eqnarray}

\noindent It reads explicitly as
(let it be  $K^2=\Lambda$)
\begin{eqnarray}
 [  ( k+p ) X- ( k-p ) Y ]\; ^{2}{\Lambda}^{2} -
 [  ( 4\,{k}^{2}{Y}^{2}-2\,{k}^{2}+2\,{p}^{2} ) {X}^{2}
 \nonumber
 \\
 + (
-4\,{k}^{2}+4\,{p}^{2} ) YX + ( -2\,{k}^{2}+2\,{p}^{2} )
{Y}^{2}+4\,{k}^{2} ]\; \Lambda +
 [ X ( k-p ) - ( k+p ) Y ] ^{2}=0\; ;
 \label{ 5.8a }
 \end{eqnarray}

\noindent the  roots are
\begin{eqnarray}
\Lambda_{1} =
{\frac{
2k\sqrt {
( YX    -   1 )^{2}
(  (  ( {Y}^{2}-1 ) {X}^{2}-{Y}^{2} +   1 ) {k}^{2}
+   {p}^{2} ( X+Y ) ^{2} ) }
}
{ (  ( k+p ) X- ( k-p ) Y ) ^{2}}
}
+
\nonumber
\\
+
{\frac
{
(2{k}^{2}{Y}^{2}    +   {p}^{2} -   {k}^{2}) {X}^{2}   + Y ({p}^{2} -{k}^{2}   ) ( 2 X  +    {Y})  +   2{k}^{2}}
{(  ( k+p ) X- ( k-p ) Y ) ^{2}}\; ,
}
\nonumber
\end{eqnarray}
\begin{eqnarray}
\Lambda_{2} =
{\frac
{
-2k\sqrt {
( YX    -   1 )^{2}
(  (  ( {Y}^{2}-1 ) {X}^{2}-{Y}^{2} +   1 ) {k}^{2}
+   {p}^{2} ( X+Y ) ^{2} ) }
}
{ (  ( k+p ) X- ( k-p ) Y ) ^{2}}
}
+
\nonumber
\\
+
{\frac
{
(2{k}^{2}{Y}^{2}    +   {p}^{2} -   {k}^{2}) {X}^{2}
 + Y ({p}^{2} -{k}^{2}   ) ( 2 X  +    {Y})  +   2{k}^{2}}
{(  ( k+p ) X- ( k-p ) Y ) ^{2}}\; .
}
\label{5.8b}
\end{eqnarray}

\noindent We can prove additional properties of these roots:
$$
a\Lambda^{2} + b \Lambda+ c =0 , \qquad
 a(\Lambda -
\Lambda_{1})(\Lambda - \Lambda_{2})=0, \qquad \Lambda_{1} \Lambda_{2} = {c \over a} \; ,
\eqno(5.8c)
$$
first checking
$
\Lambda_{1}\Lambda_{1}^{*}=1 ,$
and then with the help of Vieta's theorem we derive a needed identity
for second root
\begin{eqnarray}
 (\Lambda_{1}\Lambda_{1}^{*})(\Lambda_{2}\Lambda_{2}^{*})= {{c\; c^*}\over {a\; a^*}},
 \qquad
 \Lambda_{2}\Lambda_{2}^{*} =
{{( X ( k-p ) - ( k+p ) Y )^{2}
(
{\frac
{k-p}
{X}
}
-
{\frac
{k+p}
{Y}
}
) ^{2}}
\over
{
(  (k   +   p ) X   - ( k-p ) Y )^{2}
(
{\frac
{k+p}
{X}
}
-
{\frac
{k-p}
{Y}} )^{2}}} \qquad \Longrightarrow  \qquad
\Lambda_{2}\Lambda_{2}^{*}=1 \; .
\nonumber
\end{eqnarray}

\section{The cases of three independent phases }

There exist 6 possibilities to
determine phases with the help of 3 parameters:
\begin{eqnarray}
 \left. \begin{array}{lrrrrrr}
\rho =   & \; F & \;G & \;  F & \; F & \; F & \;F       \\
\mu =    & \; F & \;  H & \; G & \; G & \; G  & G  \\
\sigma = & \; G & \;  F & \;  F & \;-F & \; H & H    \\
\nu =    & \; H & \;  F & \;  H & \; H  & F & -F \\
\end{array} \right.  .
\label{6.1}
\end{eqnarray}

\noindent Let us consider   the variant 1  in (\ref{6.1}):
\begin{eqnarray}
\det S = \left | \begin{array}{rrrr}
 (\alpha - e^{iF})  &  (\beta  - e^{iF})   & (\alpha - e^{iF}) K  &  (\beta  - e^{iF}) K \\
 (\alpha -e^{iF} )K  &    (\beta-e^{iF}) K & (\alpha -e^{iF}) & ( \beta  -e^{iF})  \\
(\alpha -e^{iG})  (k +p)  & (\beta -e^{iG}) (k-p)  & -
   (\alpha-e^{iG}) (k-p )K & -  (\beta -e^{iG})( k+p)K \\
    (\alpha -e^{iH}) (k +p)K  &   (\beta -e^{iH})(k-p )K &  -
  (\alpha -e^{iH}) (k-p)  & - ( \beta -e^{iH}) (k+p )
  \end{array} \right |.
 \label{6.2a}
\end{eqnarray}

\noindent With notation
$
e^{iF} = x, \; e^{G} = y, \; e^{iH} = z\; ;
$
it reads
\begin{eqnarray}
\det S = \left | \begin{array}{rrrr}
 (\alpha - x)  &  (1 - \alpha x) \alpha^{-1}   & (\alpha - x) K  &  (1 - \alpha x) \alpha^{-1}  K \\
 (\alpha -x )K  &    (1 - \alpha x) \alpha^{-1} K & (\alpha -x) & (1 - \alpha x) \alpha^{-1}  \\
(\alpha -y)  (k +p)  & (1 - \alpha y)\alpha^{-1}  (k-p)  & -
   (\alpha-y) (k-p )K & -  (1 - \alpha y)\alpha^{-1} ( k+p)K \\
    (\alpha -z) (k +p)K  &   (1 - \alpha z) \alpha^{-1} (k-p )K &  -
  (\alpha -z) (k-p)  & - (1 - \alpha z) \alpha^{-1} (k+p )
  \end{array} \right |
\nonumber
\end{eqnarray}
\begin{eqnarray}
= (1 - \alpha x) \alpha^{-1} (1 - \alpha x) \alpha^{-1} (1 -
\alpha y)\alpha^{-1} (1 - \alpha z)\alpha^{-1} \alpha^{2}
\nonumber
\\
\times \left | \begin{array}{rrrr}
 X  &  1   & XK  &   K \\
  X K  &     K &   X  & 1 \\
Y  (k +p)  &   (k-p)  & -   Y (k-p )K & -  ( k+p)K \\
     \Lambda (k +p)K  &    (k-p )K &  -   \Lambda (k-p)  & - (k+p )
  \end{array} \right |.
\nonumber
\end{eqnarray}

\noindent So we arrive at
\begin{eqnarray}
\det \left | \begin{array}{rrrr}
 X  &  1   & XK  &   K \\
  X K  &     K &   X  & 1 \\
Y  (k +p)  &   (k-p)  & -   Y (k-p )K & -  ( k+p)K \\
     Z (k +p)K  &    (k-p )K &  -   Z (k-p)  & - (k+p )
  \end{array} \right | =0\; ,
 \label{6.2b}
 \end{eqnarray}

\noindent which results in a bi-quadratic equation (let $\Lambda =K^{2}$)
\begin{eqnarray}
[  ( k+p ) X- ( k-p ) Y ] [ ( k-p ) X- Z ( k+p )  ] \Lambda^{2}
\nonumber
\\
+ 2[ X( {k}^{2}   +   {p}^{2} )  ( Y+Z ) - ({k}^{2}-{p}^{2} )  (
{X}^{2}+YZ )  ] \Lambda
\nonumber
\\
+ [  ( k-p ) X- ( k+p ) Y ]
 [ ( k+p ) X-Z ( k-p )  ] =0 \; ;
\label{6.2c}
\end{eqnarray}

\noindent
its roots are
\begin{eqnarray}
\Lambda_{1} =1,\quad
\Lambda_{2}  = {  [   ( k-p ) X - ( k+p ) Y ]\; [  ( k+p ) X - \Lambda ( k-p ) ]\;
\over   [  ( k+p ) X - ( k-p ) Y ] \;  [ ( k-p ) X - Z ( k+p ) ]  } , \qquad
\Lambda_{2} \; \Lambda^{*}_{2} = 1 \; .
\label{6.2d}
\end{eqnarray}

Consider  the variant  2 in  (\ref{6.1}):
\begin{eqnarray}
\det S = \left | \begin{array}{rrrr}
 (\alpha - e^{iG})  &  (\beta  - e^{iG})   & (\alpha - e^{iG}) K  &  (\beta  - e^{iG}) K \\
 (\alpha -e^{iH} )K  &    (\beta-e^{iH}) K & (\alpha -e^{iH}) & ( \beta  -e^{iH})  \\
(\alpha -e^{iF})  (k +p)  & (\beta -e^{iF}) (k-p)  & -
   (\alpha-e^{iF}) (k-p )K & -  (\beta -e^{iF})( k+p)K \\
    (\alpha -e^{iF}) (k +p)K  &   (\beta -e^{iF})(k-p )K &  -
  (\alpha -e^{iF}) (k-p)  & - ( \beta -e^{iF}) (k+p )
  \end{array} \right |,
 \label{6.3a}
 \end{eqnarray}
in notation
$
e^{iF} = x, \;  e^{G} = y, \;  e^{iH} = z\; ,
$ it reads
$$
\det S
$$
\begin{eqnarray}
\hspace{-15mm} =\left | \begin{array}{rrrr}
 (\alpha - y)  &  ( 1 - \alpha y) \alpha^{-1}    & (\alpha - y) K  &  (1 - \alpha y) \alpha^{-1}  K \\
 (\alpha -z )K  &    (1 - \alpha z)\alpha^{-1}  K & (\alpha -z) & ( 1 - \alpha z)\alpha^{-1}   \\
(\alpha -x)  (k +p)  & (1 - \alpha x)\alpha^{-1}  (k-p)  & -
   (\alpha-x) (k-p )K & -  (1 - \alpha x )\alpha^{-1} ( k+p)K \\
    (\alpha -x) (k +p)K  &   (1 - \alpha x)\alpha^{-1} (k-p )K &  -
  (\alpha -x) (k-p)  & - ( 1 - \alpha x) \alpha^{-1} (k+p )
  \end{array} \right |
\nonumber
\end{eqnarray}
\begin{eqnarray}
= ( 1 - \alpha y) \alpha^{-1} (1 - \alpha z)\alpha^{-1} (1 -
\alpha x)^{2}\alpha^{-2} \; \alpha^{2}
\nonumber
\\
\times \left | \begin{array}{rrrr}
  Y  &  1     &  Y K  &    K \\
  Z K  &     K &  Z  & 1    \\
 X (k +p)  &   (k-p)  & -    X (k-p )K & -   ( k+p)K \\
     X  (k +p)K  &    (k-p )K &  -
  X (k-p)  & - (k+p )
  \end{array} \right |;
\nonumber
\end{eqnarray}

\noindent
thus we arrive at
\begin{eqnarray}
\det \left | \begin{array}{rrrr}
  Y  &  1     &  Y K  &    K \\
  Z K  &     K &  Z  & 1    \\
 X (k +p)  &   (k-p)  & -    X (k-p )K & -   ( k+p)K \\
     X  (k +p)K  &    (k-p )K &  -
  X (k-p)  & - (k+p )
  \end{array} \right | =0 \; .
\label{6.3b}
\end{eqnarray}

\noindent
Explicitly, we have a bi-quadratic equation
 (let $\Lambda =K^{2}$)
\begin{eqnarray}
[    ( k+p  ) X - Z  ( k-p  )  ]\;
  [  ( k-p  ) X-Y  ( k+p  )    ] \Lambda ^{2} +
\nonumber
\\
 +2  [ (k ^{2}+ p ^{2}  )   ( Y+Z  ) X
 -  ( k^{2}- p^{2}  )   (  X ^{2}+YZ  )      ]\;
\Lambda
 +
\nonumber
\\
 +
 [   ( k+p  ) X-Y  ( k-p  )  ]\;
 [  ( k-p  ) X-Z  ( k+p  )   ]\;
  =0 ;
\label{6.3c}
\end{eqnarray}

\noindent
its roots are
\begin{eqnarray}
\Lambda _{1}=1,\quad
\Lambda _{2 }= {\frac { [ ( k-p )X  -  ( k+p  ) Z ]  \; [ ( k+p) X  -  ( k-p  ) Y ]   }
{  [ ( k+p) X -  ( k-p  ) Z ]  \; [ ( k-p) X -  ( k+p  ) Y ]  }} \; , \qquad
\Lambda _{2} \; \Lambda ^{*}_{2} = 1 \; .
\label{6.3d}
\end{eqnarray}

Consider the variant  3  in (\ref{6.1}):
\begin{eqnarray}
\det S = \left | \begin{array}{rrrr}
 (\alpha - e^{iF})  &  (\beta  - e^{iF})   & (\alpha - e^{iF}) K  &  (\beta  - e^{iF}) K \\
 (\alpha -e^{iG} )K  &    (\beta-e^{iG}) K & (\alpha -e^{iG}) & ( \beta  -e^{iG})  \\
(\alpha -e^{iF})  (k +p)  & (\beta -e^{iF}) (k-p)  & -
   (\alpha-e^{iF}) (k-p )K & -  (\beta -e^{iF})( k+p)K \\
    (\alpha -e^{iH}) (k +p)K  &   (\beta -e^{iH})(k-p )K &  -
  (\alpha -e^{iH}) (k-p)  & - ( \beta -e^{iH}) (k+p )   \end{array} \right |;
 \nonumber
 \label{6.4a}
 \end{eqnarray}
 in notation
 $
e^{iF} = x, \; e^{G} = y, \; e^{iH} = z\; ;
$ it reads
$$
\det S = \left | \begin{array}{rrrr}
 (\alpha - e^{iF})  &  (\beta  - e^{iF})   & (\alpha - e^{iF}) K  &  (\beta  - e^{iF}) K \\
 (\alpha -e^{iG} )K  &    (\beta-e^{iG}) K & (\alpha -e^{iG}) & ( \beta  -e^{iG})  \\
(\alpha -e^{iF})  (k +p)  & (\beta -e^{iF}) (k-p)  & -
   (\alpha-e^{iF}) (k-p )K & -  (\beta -e^{iF})( k+p)K \\
    (\alpha -e^{iH}) (k +p)K  &   (\beta -e^{iH})(k-p )K &  -
  (\alpha -e^{iH}) (k-p)  & - ( \beta -e^{iH}) (k+p )   \end{array} \right |
   $$
   $$
= \left | \begin{array}{rrrr}
 (\alpha - x)  &  (1 - \alpha x) \alpha^{-1}   & (\alpha - x) K  &  (1 - \alpha x )\alpha^{-1} K \\
 (\alpha -y )K  &    (1 - \alpha y) \alpha^{-1} K & (\alpha -y) & ( 1 - \alpha y) \alpha^{-1} \\
(\alpha - x)  (k +p)  & (1 -\alpha x)\alpha^{-1} (k-p)  & -
   (\alpha-x) (k-p )K & -  (1 - \alpha x ) \alpha^{-1} ( k+p)K \\
    (\alpha -z) (k +p)K  &   (1 - \alpha z) \alpha^{-1} (k-p )K &  -
  (\alpha -z) (k-p)  & - ( 1 - \alpha z )\alpha^{-1} (k+p )   \end{array} \right |
$$
\begin{eqnarray}
=(1 - \alpha x) \alpha^{-1} (1 - \alpha y) \alpha^{-1}(1 - \alpha
x) \alpha^{-1}( 1 - \alpha z )\alpha^{-1}\alpha^{2}
\nonumber
\\
\times
 \left | \begin{array}{rrrr}
 X &  1    &  X K  &   K \\
  Y K  &     K &  Y & 1  \\
 X   (k +p)  &  (k-p)  & -
   X (k-p )K & -   ( k+p)K \\
    Z  (k +p)K  &   (k-p )K &  -
   Z  (k-p)  & -  (k+p )   \end{array} \right |= 0,
\label{6.4b}
\end{eqnarray}

\noindent
which gives a bi-quadratic equation
 (let $\Lambda =K^{2}$)
\begin{eqnarray}
pX [ k ( Y  -   Z ) - ( Y   +   Z ) p ]\; \Lambda^2
\nonumber
\\
+   2[  ( X^2  +   YZ ) k^2 - ( k^2    -   p^2 )  ( Y+Z ) X]\;
\Lambda -
   pX [ k ( Y  -   Z ) + ( Y   +   Z ) p ] =0 \; ; \label{6.4c}
\end{eqnarray}

\noindent
its roots are
\begin{eqnarray}
\Lambda_{1}= {\frac {\sqrt { k^2 [ ( X-Z )^2 ( X-Y ) ^2 k^2
 + 2 p^2 (  ( Y+Z ) X-2YZ ) X ( -1/2Z-1/2Y+X )  ] }
} {pX [ ( Y-Z )k  - ( Y+Z ) p ] } }
\nonumber
\\
- {\frac {  k^2 (X^2 + YZ) - (Y+Z ) (   k^2 -  p^2) X } {pX (  (
Y-Z )k - ( Y+Z ) p )}
     }\; ,
\nonumber
\\
 \Lambda_{2}=
{\frac {-\sqrt { k^2 [  ( X - Z )^2 ( X - Y )^2 k^2 + 2p^2 (  ( Y
+ Z ) X - 2YZ ) X ( - 1/2 Z - 1/2 Y + X )  ] } }
 {pX ( ( Y - Z ) k - ( Y + Z ) p ) }
}
\nonumber
\\
-{\frac { k^2 (X^2  + YZ) - ( Y + Z )( k^2 - p^2) X} {pX [ ( Y - Z
) k - ( Y + Z ) p ]} }\; .
\label{6.4d}
\end{eqnarray}

\noindent
By direct calculation we get
$
\Lambda_{1}\Lambda_{1}^*=1 \; ;
$
then with the help of Vieta's  theorem
we prove the same for second root
$$
(\Lambda_{1}\Lambda_{1}^{*})(\Lambda_{2}\Lambda_{2}^{*})= {{c\; c^*}\over {a\; a^*}}\;,
$$
$$
\Lambda_{2}\Lambda_{2}^{*}=
{( k ( Y-Z ) + ( Y+Z ) p )  ( k ( {Y}^{-1}-{Z}^{-1} ) + ( {Y}^{-1}+{Z}^{-1} ) p )
\over
{
 ( k ( Y-Z ) - ( Y+Z ) p )  ( k ( {Y}^{-1}-{Z}^{-1} ) - ( {Y}^{-1}+{Z}^{
-1} ) p )
}}
\qquad \Longrightarrow  \qquad
\Lambda_{2}\Lambda_{2}^{*}=1.
$$

Consider  the variant 4  in  (\ref{6.1}):
\begin{eqnarray}
\det S = \left | \begin{array}{rrrr}
 (\alpha - e^{iF})  &  (\beta  - e^{iF})   & (\alpha - e^{iF}) K  &  (\beta  - e^{iF}) K \\
 (\alpha -e^{iG} )K  &    (\beta-e^{iG}) K & (\alpha -e^{iG}) & ( \beta  -e^{iG})  \\
(\alpha -e^{-iF})  (k +p)  & (\beta -e^{-iF}) (k-p)  & -
   (\alpha-e^{-iF}) (k-p )K & -  (\beta -e^{-iF})( k+p)K \\
    (\alpha -e^{iH}) (k +p)K  &   (\beta -e^{iH})(k-p )K &  -
  (\alpha -e^{iH}) (k-p)  & - ( \beta -e^{iH}) (k+p )   \end{array} \right |.
 \label{6.5a}
 \end{eqnarray}
 from whence with the use of notation $e^{iF} = x, \;e^{G} = y, \; e^{iH} = z\; ;
$
 we get
$$
\det S
$$
$$
\hspace{-5mm} = \left | \begin{array}{rrrr}
 (\alpha - x)  &  (1 - \alpha x) \alpha^{-1}   & (\alpha - x) K  &  (1 - \alpha x )\alpha^{-1} K \\
 (\alpha -y )K  &    (1 - \alpha y) \alpha^{-1} K & (\alpha -y) & ( 1 - \alpha y) \alpha^{-1} \\
- (1 - \alpha  x) x^{-1}  (k +p)  & - (\alpha -x )\alpha^{-1}
x^{-1}  (k-p)  &
   (1 -\alpha x) (k-p )x^{-1} K &  (\alpha -x )\alpha^{-1} x^{-1}( k+p)K \\
    (\alpha -z) (k +p)K  &   (1 - \alpha z) \alpha^{-1} (k-p )K &  -
  (\alpha -z) (k-p)  & - ( 1 - \alpha z )\alpha^{-1} (k+p )   \end{array} \right |
   $$
\begin{eqnarray}
=(1 - \alpha x) \alpha^{-1} (1 - \alpha y) \alpha^{-1} (1 - \alpha
x) \alpha^{-1} x^{-1} (1 - \alpha z) \alpha^{-1}\;
\alpha^{2}
\nonumber
\\
\times \left | \begin{array}{rrrr}
 X  &  1  &  X K  &   K \\
 Y K  &     K &  Y  & 1 \\
-   (k +p)  & - X  (k-p)  &
    (k-p ) K &   X ( k+p)K \\
     Z (k +p)K  &    (k-p )K &  -
  Z (k-p)  & -  (k+p )   \end{array} \right |;
 \label{6.5b}
 \end{eqnarray}

\noindent
equation  $\det S =0$  is a bi-quadratic one (let $\Lambda =K^{2}$)
\begin{eqnarray}
  [   (X^{2} - 1)k + p(X^{2} + 1)  ]
[( Y-Z)k - p(Y + Z)]\Lambda^{2}
\nonumber
\\
- \left \{ [      2( Y +Z )X^{2}  - 4(YZ +1)X + 2(Y + Z)  ]k^{2} -
2p^{2}(X^{2} + 1)(Y + Z) \right \} \Lambda
\nonumber
\\
+ [(X^{2} - 1)k - (X^{2}+1)p]\; [( Y-Z  )k + p(Y+Z)]=0\; ;
\label{6.5c}
\end{eqnarray}
its roots are
$$
 \Lambda_{1}
 $$
 \begin{eqnarray}
 = {\frac
{2k\sqrt {  ( XZ-1 )  ( X-Z )  ( XY-1 )
 ( X-Y ) k^2 - [ YZ(X^2  +1) -  X ( Y + Z )  ]
 [ 1 + X^2 -  X( Y + Z ) ]p^2}
} { [  ( X^2 - 1 ) k + p ( X^2 + 1 )  ][    ( Y-Z ) k-p ( Y+Z ) ])
}
}
 \nonumber
\\
+ {\frac { [  ( Y+Z ) {X}^{2} -2( YZ+1 ) X+Y+Z ] k^2  -  p^2(Y +
Z) (X^2 +1)} {[  ( X^2 - 1 ) k + p ( X^2 + 1 )  ][  ( Y-Z ) k-p
( Y+Z )  ]} }
\nonumber
\end{eqnarray}
$$
 \Lambda_{2}
$$
=\begin{eqnarray}{\frac {-2k\sqrt {  ( XZ-1 )  ( X-Z )  ( XY-1 )
 ( X-Y ) k^2 - [ YZ(X^2  +1) -  X ( Y + Z )  ]
 [ 1 + X^2 -  X( Y + Z ) ]p^2}
} { [  ( X^2 - 1 ) k + p ( X^2 + 1 ) ][  ( Y-Z ) k-p ( Y+Z )  ]
} }
\nonumber
\\
+{\frac { [  ( Y+Z ) {X}^{2} -2( YZ+1 ) X+Y+Z ] k^2  -  p^2(Y + Z)
(X^2 +1)} { [  ( X^2 - 1 ) k + p ( X^2 + 1 )  ][  ( Y - Z ) k -
p ( Y + Z )  ] } }.
\nonumber
\label{6.5d}
\end{eqnarray}

\noindent
The first root is a phase
$
\Lambda_{1}\Lambda_{1}^*=1,
$
with the use of Vieta's  theorem we prove the same for second root:
$$
(\Lambda_{1}\Lambda_{1}^{*})(\Lambda_{2}\Lambda_{2}^{*})= {{c\; c^*}\over {a\; a^*}} \;, \qquad
\Lambda_{2}\Lambda_{2}^{*}
$$
\begin{eqnarray}
=
{{(  ( {X}^{2}-1 ) k-p ( {X}^{2}+1 )
 )
( k ( Y-Z ) + ( Y+Z ) p )
(  ( {{1}\over { X^2 }}-1 ) k-p ({{1}\over { X^2 }}+1 )
 )  ( k ({{1}\over { Y }}-{{1}\over { Z }} ) + ( {{1}\over {Y  }}+
{{1}\over { Z }} ) p )
}
\over
{
  (  ( {X}^{2}-1 ) k+p
 ( {X}^{2}+1 )  ) ( k ( Y-Z ) -
 ( Y+Z ) p ) (  ( {{1}\over { X^2 }}-1 )
k+p ( {{1}\over {X^2  }}+1 )  ) ( k ( {{1}\over { Y }}-
{{1}\over { Z }}) - ( {{1}\over {Y}}+  {{1}\over { Z }} ) p ) }} =1\;.
\nonumber
\end{eqnarray}

Consider  the variant  5 in (\ref{6.1}):
\begin{eqnarray}
\det S= \left | \begin{array}{rrrr}
 (\alpha - e^{iF})  &  (\beta  - e^{iF})   & (\alpha - e^{iF}) K  &  (\beta  - e^{iF}) K \\
 (\alpha -e^{iG} )K  &    (\beta-e^{iG}) K & (\alpha -e^{iG}) & ( \beta  -e^{iG})  \\
(\alpha -e^{iH})  (k +p)  & (\beta -e^{iH}) (k-p)  & -
   (\alpha-e^{iHF}) (k-p )K & -  (\beta -e^{iHF})( k+p)K \\
    (\alpha -e^{iF}) (k +p)K  &   (\beta -e^{iF})(k-p )K &  -
  (\alpha -e^{iF}) (k-p)  & - ( \beta -e^{iF}) (k+p )   \end{array} \right |;
 \label{6.6a}
 \end{eqnarray}
 with the notation $e^{iF} = x, \; e^{G} = y, \; e^{iH} = z$,
it reads
$$
\det S
$$
$$
= \left | \begin{array}{rrrr}
 (\alpha - x)  &  (1 - \alpha x ) \alpha^{-1}   & (\alpha - x) K  &  (1 - \alpha x) \alpha^{-1} K \\
 (\alpha -y )K  &    (1 - \alpha y) \alpha^{-1} K & (\alpha -y) & ( 1- \alpha y) \alpha^{-1}  \\
(\alpha -z)  (k +p)  & (1 - \alpha z)\alpha^{-1} (k-p)  & -
   (\alpha- z) (k-p )K & -  (1 -\alpha z )\alpha^{-1} ( k+p)K \\
    (\alpha -x) (k +p)K  &   (1 - \alpha x) \alpha^{-1} (k-p )K &  -
  (\alpha -x ) (k-p)  & - ( 1 -\alpha x ) \alpha^{-1} (k+p )   \end{array} \right |
  $$
$$
= (1 - \alpha x ) \alpha^{-1} (1 - \alpha y) \alpha^{-1} (1 -
\alpha z)\alpha^{-1} (1 - \alpha x) \alpha^{-1}\; \alpha^{2}\times
$$
\begin{eqnarray}
\times \left | \begin{array}{rrrr}
  X  &  1   &  X  K  &  K \\
  Y K  &    K &  Y  & 1  \\
 Z  (k +p)  &  (k-p)  & -     Z  (k-p )K & -  ( k+p)K \\
    X (k +p)K  &    (k-p )K &  -    X  (k-p)  & -  (k+p )   \end{array} \right |;
 \label{6.6b}
 \end{eqnarray}

\noindent
 explicit form of $\det S=0$  is (let $\Lambda = K^{2}$)
\begin{eqnarray}
 \left \{   [   ( k+p  )X - Y( k-p  )   ]K^{2} + X ( k-p) -  Y( k+p) \right \}
\nonumber
\\
\times \left \{       [ (k+p) X - Z (k - p) ]K^{2} + X( k - p)  -  Z( k+p
) \right \} =0
 \label{6.6b}
 \end{eqnarray}

\noindent
the roots are
\begin{eqnarray}
\Lambda_{1} =-\frac { X  ( k-p  ) -Z
( k+p  )    } {( k+p  ) X-Z  ( k-p  ) }\; ,\qquad \Lambda_{2}=
-\frac { X( k-p  ) - Z  (
k+p  )  } {  ( k+p  ) X - Z  ( k-p  ) } \; ;
 \label{6.6c}
 \end{eqnarray}
 identities hold
$
\Lambda_{1} \Lambda_{1}^{*}=1, \quad   \Lambda_{2} \Lambda_{2}^{*}=1\;  .
 $

Consider   the variant  6 in (\ref{6.1}):
\begin{eqnarray}
\det S = \left | \begin{array}{rrrr}
 (\alpha - e^{iF})  &  (\beta  - e^{iF})   & (\alpha - e^{iF}) K  &  (\beta  - e^{iF}) K \\
 (\alpha -e^{iG} )K  &    (\beta-e^{iG}) K & (\alpha -e^{iG}) & ( \beta  -e^{iG})  \\
(\alpha -e^{iH})  (k +p)  & (\beta -e^{iH}) (k-p)  & -
   (\alpha-e^{iH}) (k-p )K & -  (\beta -e^{iH})( k+p)K \\
    (\alpha -e^{-iF}) (k +p)K  &   (\beta -e^{-iF})(k-p )K &  -
  (\alpha -e^{-iF}) (k-p)  & - ( \beta -e^{-iF}) (k+p )   \end{array} \right |,
  \label{6.7a}
  \end{eqnarray}
with notation $
e^{iF} = x, \; e^{G} = y, \; e^{iH} = z$ it can be presented as
$$
\det S
$$
$$
= \left | \begin{array}{rrrr}
 (\alpha - x)  &  (1 - \alpha x )\alpha^{-1}    & (\alpha - x) K  &  (1 - \alpha x)\alpha^{-1}  K \\
 (\alpha -y )K  &    ( 1 -\alpha y)\alpha^{-1}  K & (\alpha -y) & ( 1 - \alpha y )\alpha^{-1}   \\
(\alpha -z)  (k +p)  & (1 - \alpha z )\alpha^{-1}  (k-p)  & -
   (\alpha-z) (k-p )K & -  (1 - \alpha z ) \alpha^{-1} ( k+p)K \\
    -(1 - \alpha x) x^{-1}  (k +p)K  &   -(\alpha - x)\alpha^{-1} x^{-1} (k-p )K &
   (1 - \alpha x )x^{-1}  (k-p)  & (\alpha - x)\alpha^{-1} x^{-1} (k+p )   \end{array} \right |
  $$
\begin{eqnarray}
= (1 - \alpha )\alpha^{-1} ( 1 -\alpha y)\alpha^{-1} (1 - \alpha z
)\alpha^{-1} (1 - \alpha x) x^{-1} \alpha^{-1}\; \alpha^{2}
\nonumber
\\
\times \left | \begin{array}{rrrr}
X  &  1   &  X  K  &   K \\
 Y K  &      K & Y & 1   \\
 Z  (k +p)  &     (k-p)  & -    Z (k-p )K & -  ( k+p)K \\
    - (k +p)K  &   - X (k-p )K &   (k-p)  &  X  (k+p )   \end{array} \right |;
 \label{6.7b}
 \end{eqnarray}

\noindent
  equation $\det S =0$ is a bi-quadratic one (let $\Lambda=K^{2}$):
\begin{eqnarray}
[-( {k}^{2}-{p}^{2} ) (Y{X}^{2} +Z) +( (k-p)^{2})ZY+(k+p)^{2})X ]
\Lambda^{2}
\nonumber
\\
+ [ ( 2( k^2-p^2)Y-4Zk^2)X^2+ (2( k^{2}-p^{2})(ZY + 1)X
 - 4k^{2}Y +2( k^{2} - p^{2})Z
] \Lambda
\nonumber
\\
 - ({k} - p)^{2}( Y{X}^2 + Z )
+ (( {k}+ {p})^{2}ZY + (k-p)^2)X=0 \; ;
\label{6.7c}
\end{eqnarray}

\noindent
the roots are
$$
\Lambda_{1}
$$
\begin{eqnarray}
={\frac {2k\sqrt {- (  ( Y-Z )  ( X^2-1 )
   ( XZ-1 )  ( X-Y ) k^2
 - ( YZ(X^2  -1)+ ( Z-Y ) X ) ( {X}^{2}+ ( Z-Y ) X -1 ) p^2  ) }
} { ( Y ( k-p ) X-k-p ) ( X ( k+p ) -Z ( k-p )  ) } }
\nonumber
\\
+{\frac {(  ( Y-2Z ) k^2 - Yp^2 ) X^2
 +   ( YZ+1 ) (k^2 - p^2) X
 + ( Z-2Y ) k^2 - Zp^2 }
{ ( Y ( k-p ) X-k-p ) ( X ( k+p ) -Z ( k-p )  )} },
\nonumber
\end{eqnarray}

$$
 \Lambda_{2}
 $$
\begin{eqnarray}
=  {\frac
{-2k\sqrt {- (  ( Y-Z )  ( X^2-1 )
   ( XZ-1 )  ( X-Y ) k^2
 - ( YZ(X^2  -1)+ ( Z-Y ) X ) ( {X}^{2}+ ( Z-Y ) X -1 ) p^2  ) }
} { ( Y ( k-p ) X-k-p ) ( X ( k+p ) -Z ( k-p )  ) } }
\nonumber
\\
+{\frac {(  ( Y-2Z ) k^2 - Yp^2 ) X^2
 +   ( YZ+1 ) (k^2 - p^2) X
 + ( Z-2Y ) k^2 - Zp^2 }
{ ( Y ( k-p ) X-k-p ) ( X ( k+p ) -Z ( k-p )  )} }\; .
\nonumber
\end{eqnarray}

The root $\Lambda_{1}$ is a phase , $
\Lambda_{1}\Lambda_{1}^{*}=1$;
then with the help of Vieta's theorem we prove the same for the second root
$$
(\Lambda_{1}\Lambda_{1}^{*})(\Lambda_{2}\Lambda_{2}^{*})= {{c\;
c^*}\over {a\; a^*}} \; ,
$$
$$
\Lambda_{2}\Lambda_{2}^{*}=
 {{( X ( k-p ) -Z ( k+p )  )  ( YX ( k+p ) -k+p )  ( {\frac {k+p}{YX}}-k+p )  ( {\frac {k-p}{X}}-{\frac {k+p}{Z}} ) }
 \over
{  ( Y ( k-p ) X-k-p )  ( X ( k+p ) -Z ( k-p )  )   ( {\frac
{k-p}{YX}}-k-p )   ( {\frac {k+p}{X}}-{\frac {k-p}{Z}} )  }} =0  \; .
$$

\section{Analysis of the general case of four phases  }

Let us turn to equations (\ref{3.7}) in general case of all four independent phases:
\begin{eqnarray}
\left | \begin{array}{rrrr}
 (\alpha - e^{i\rho})  &  (\beta  - e^{i\rho})   & (\alpha - e^{i\rho}) K  &  (\beta  - e^{i\rho}) K \\
 (\alpha -e^{i\mu} )K  &    (\beta-e^{i\mu}) K & (\alpha -e^{i\mu}) & ( \beta  -e^{i\mu})  \\
(\alpha -e^{i\sigma})  (k +p)  & (\beta -e^{i\sigma}) (k-p)  & -
   (\alpha-e^{i\sigma}) (k-p )K & -  (\beta -e^{i\sigma})( k+p)K \\
    (\alpha -e^{i\nu}) (k +p)K  &   (\beta -e^{i\nu})(k-p )K &  -
  (\alpha -e^{i\nu}) (k-p)  & - ( \beta -e^{i\nu}) (k+p )
  \end{array} \right |.
 \label{7.1}
 \end{eqnarray}

Below, the shortening notation will be used:
\begin{eqnarray}
e^{i\rho} =x, \qquad (\alpha - e^{i\rho}) = \alpha - x, \; (\beta
- e^{i\rho})=(1 -\alpha x) / \alpha, \; X = {\alpha - x \over 1
-\alpha x}, \;X^{*}={1 \over X}\; ,
\nonumber
\\
e^{i\mu} =y , \qquad (\alpha - e^{i\mu}) = \alpha - y, \;  (\beta
- e^{i\mu})=(1 -\alpha y) / \alpha, \; Y = {\alpha - y \over 1
-\alpha y}, \;Y^{*}={1 \over Y}\; ,
\nonumber
\\
e^{i\sigma} =v, \qquad (\alpha - e^{i\sigma}) = \alpha - v, \;
(\beta  - e^{i\sigma})=(1 -\alpha v) / \alpha, \; V = {\alpha - v
\over 1 -\alpha v}, \;V^{*}={1 \over V}\; ,
\nonumber
\\
e^{i\nu} =w, \qquad (\alpha - e^{i\nu}) = \alpha - w, \; (\beta  -
e^{i\nu})=(1 -\alpha w) / \alpha, \; W {\alpha - w \over 1 -\alpha
w}, \;W^{*}={1 \over W}\; .
\label{7.2}
\end{eqnarray}

Equation $\det S=0$ can be presented as follows:
\begin{eqnarray}
\det S ={ (1 -\alpha  x) \over \alpha} { (1 - \alpha  y) \over
\alpha} { (1 - \alpha  v) \over \alpha}  { (1 -\alpha  w) \over
\alpha}\;\alpha^{2}
\nonumber
\\
\times \left | \begin{array}{rrrr}
   X   &    1 &   \alpha  XK  &  K \\
 Y K  &    K & \alpha Y  & 1 \\
  (k +p) V   & (k-p)  & -
    (k-p )  V K & -  ( k+p)K \\
 (k +p) W K  & (k-p)K   & -     (k-p )  W & -  ( k+p)
  \end{array} \right |,
 \label{7.3}
 \end{eqnarray}

\noindent which results in a bi-quadratic equation
 (let  $\Lambda
=K^{2}$)
\begin{eqnarray}
[   ( V-X  )k - p( V+X  )   ][  ( W-Y  )k + p( W+Y  )   ]\;
\Lambda^{2}
\nonumber
\\
+2\{  [ ( 2X - W - Y  )V -( X -2  Y  )W - XY ] k^{2} + p^{2}( W+Y
)( V + X  ) \}  \Lambda
\nonumber
\\
+ [   (W - Y  )k - p( W+Y ) ] [   (V - X )k + p(V + X)]=0 \; .
\label{7.4}
\end{eqnarray}

\noindent
The roots  are
$$
\Lambda_{1}=
$$
\begin{eqnarray}{\frac {2k\sqrt {
 [ ( ( X-Y ) W+XY ) V - WXY ] ( V+X-Y-W )  p^2
- ( X-Y ) ( W-X ) ( V-Y ) ( V-W ) k^2
         }
} { [ ( W-Y ) k + ( W+Y )p  ] [ ( V-X ) k - ( V+X )p  ] } }
\nonumber
\\
+ {\frac { [( W-2X+Y ) V +( X-2Y ) W + XY ] k^2 - ( W +Y )  (V + X
) p^2 } { [( W-Y ) k  + ( W+Y )p  ] [ ( V-X ) k - ( V+X )p  ]  }
},\hspace{20mm}
\nonumber
\label{7.5a}
\end{eqnarray}
$$
\Lambda_{2}=
$$
\begin{eqnarray}
{\frac {-2k\sqrt { [ ( ( X-Y ) W+XY ) V - WXY ] ( V+X-Y-W )  p^2 -
( X-Y ) ( W-X ) ( V-Y ) ( V-W ) k^2
         }
} { [ ( W-Y ) k + ( W+Y )p  ] [ ( V-X ) k - ( V+X )p  ] } }
\nonumber
\\
+ {\frac { [( W-2X+Y ) V +( X-2Y ) W + XY ] k^2 - ( W +Y )  (V + X
) p^2 } {[ ( W-Y ) k  + ( W+Y )p  ] [ ( V-X ) k - ( V+X )p  ] } }.\hspace{20mm}
\nonumber
\label{7.5b}
\end{eqnarray}

\noindent
We readily prove that both  complex roots have a unit length:
$$
\Lambda_{1}\Lambda_{1}^{*}=1 \; ,\qquad
(\Lambda_{1}\Lambda_{1}^{*})(\Lambda_{2}\Lambda_{2}^{*})= {{c\;
c^*}\over {a\; a^*}} \; ,
$$
$$
\Lambda_{2} \Lambda_{2}^{*} =
{ [ (W-Y ) k-p (W+Y )  ] \; [ (V-X ) k+p (V+X )  ] \over
[ (V-X ) k-p (V+X ) ]\; [ (W-Y ) k+p (W+Y ) ]\; } \times
$$
\begin{eqnarray}
\times { [ ({1\over W}-{1\over Y} ) k-p ({1\over
W}+{1\over Y} )  ] \; [ ({1\over V}-{1\over X} ) k+p
({1\over V}+{1\over X} )  ]  \over  [ ( {1\over V} -{1\over X } )k-p
({1\over V}+{1\over X} )  ] \; [ ({1\over W}-{1\over
Y} ) k+p ({1\over W}+{1\over Y} )  ] } =1 \; .
\label{7.6}
\end{eqnarray}

\section{Conclusion}

Thus, all solutions of the 4-th order polynomial have been found  in explicit form.
All of them  are complex numbers of unit length, and thereby they are appropriate to
produce a quantization rule for
$k$.
Some of these rules are simple, and straightforwardly give needed quantization rule
of the form    $2ka = \pi n, \pi + p n, n =0\pm 1, \pm 2, ....$.
The  the most of produced expression for the roots are complex enough
and can be solved with respect to parameter $k$ only numerically.

Till now, only simplest variants of quantization
 $2ka = \pi n, \pi + p n, n =0\pm 1, \pm 2, ....$ were examined in the literature \cite{Mostepanenk-Trunov-1997}
 in the context
 of treating the  Casimir effect for Dirac particle in external magnetic field.

A last remark should be added: the final algebraic problem in the form of polynomial and
 corresponding quantization for $k$
preserve its form when considering the problem of electron in magnetic field on the base of cylindric
 coordinates as well. Moreover, the same polynomial
arises in considering a free electron in the domain between two planes.

This  work was   supported   by the Fund for Basic Researches of Belarus,
 F 13K-079, within the cooperation framework between Belarus  and Ukraine.
O.V. Veko is grateful to the Ministry of Education of Republic of Belarus
 for the   grant for  work on probation in N.N. Bogolyubov Institute for Theoretical Physics,
National Academy of Sciences of Ukraine.
This research has been partially supported by the Armenian State Committee
of Science (SCS Grant No. 13RB-052) and  has been partially conducted
within the scope of the International Associated Laboratory (CNRS-France \& SCS-Armenia) IRMAS.
The work by A. Ishkhanyan has received funding from the European
 Union Seventh Framework Programme (FP7/2007-2013, grant No. 205025 - IPERA)
 and from the Armenian State Committee of Science (SCS Grant No. 13RB-052).

\end{document}